\documentclass[pre,twocolumn,showpacs,eqsecnum]{revtex4}
\usepackage{graphicx}
\usepackage{bm}
\usepackage{amsmath,amssymb, upgreek}

\begin{document}
\title{Energy dissipation of rigid dipoles in a viscous fluid under the action of a time-periodic field: the influence of thermal bath and dipole interaction}
\author{T.~V.~Lyutyy}
\email{lyutyy@oeph.sumdu.edu.ua}
\author{V.~V.~Reva}
\affiliation{Sumy State University, 2 Rimsky-Korsakov Street, UA-40007 Sumy, Ukraine}

\begin{abstract}
Ferrofluid heating by an external alternating field is studied based on the rigid dipole model, where the magnetization of each particle in a fluid is supposed to be firmly fixed in the crystal lattice. Equations of motion, employing the Newton's second law for rotational motion, the condition of rigid body rotation, and the assumption that the friction torque is proportional to angular velocity, are used. This oversimplification permits us to expand the model easily: to take into account the thermal noise and inter-particle interaction that allows to estimate from unified positions the role of thermal activation and dipole interaction in the heating process. Our studies are conducted in three stages. The exact expressions for the average power loss of a single particle are obtained within the dynamical approximation. Then, in the stochastic case, the power loss of a single particle is estimated analytically using the Fokker-Planck equation and numerically using the effective Langevin equation. Finally, the power loss for the particle ensemble is obtained using the molecular dynamics method. Here, the local dipole fields are calculated approximately based on the Barnes-Hut algorithm. The revealed trends in the behaviour of both a single particle and the particle ensemble suggest the way of choosing the conditions for obtaining the maximum heating efficiency. The competitiveness character of the inter-particle interaction and thermal noise is investigated in details. Two situations, when the thermal noise rectifies the power loss reduction caused by the interaction, are described. The first of them is related to the complete destruction of dense clusters at high noise intensity. The second one is originated from the rare switching of the particles in clusters due to thermal activation, when the noise intensity is relatively weak. In this way, the constructive role of noise appears in the system.
\end{abstract}
\maketitle

\section{INTRODUCTION}
\label{Int}

The ferrofluid \cite{Rosensweig1985Ferrohydrodynamics, 0038-5670-17-2-R02} response to external fields is a key factor for its applications. Firstly, in this regard, an attention should be given to high-performance microwave absorbers designed in a complex manner, which are under intense investigation now \cite{C4TA01681E, doi:10.5185/amlett.2015.5807, C5TA02381E, Varshney2017, doi:10.1002/9781119208549.ch13}. Secondly, a special attention deserves magnetic fluid hyperthermia method for cancer treatment \cite{0022-3727-36-13-201, JORDAN1999413, andr2006magnetism}. In both examples mentioned above, the energy of a periodic external field is transformed into the thermal one due to viscous rotation of the ferrofluid particles and damped precession of the magnetization within the particles crystal lattice. The heating process is often described within the quasi-equilibrium assumption  and theory of linear response to an alternating field \cite{Rosensweig2002370}. Despite the simplicity and obviousness of this concept, its application domain is narrow. As evidenced from various experiments \cite{andr2006magnetism} including the recent \cite{0022-3727-49-29-295001, doi:10.1109/TMAG.2016.2516645, doi:10.1063/1.4974803, Arteagacardona2016636, doi:10.1021/acs.molpharmaceut.5b00866, doi:10.1021/acsnano.7b01762, doi:10.2147/IJN.S141072, doi:10.1166/sam.2017.2948, doi:10.1021/acsnano.7b01762}, strong deviations from the analytical predictions occur. In a wide sense, these differences are originated from the features of individual dynamics aroused as a consequence of coupled mechanical and magnetic motions of each particle and structure of a ferrofluid aroused as a consequence of inter-particle interaction.

The analytical description of the driven coupled motion of a magnetic particle in a fluid with its magnetic moment is an interesting and complicated problem. Since the base model equations were introduced by Cebers \cite{Cebers1975}, the consistent investigation of the particle response to periodic fields is absent up to now (see Introduction in Ref. \cite{Lyutyy201887} for details). Because of computational difficulties, the simplified approaches are widely used to these purposes. The first approach is the fixed particle model \cite{PhysRevE.86.061404, PhysRevE.93.012607}, within which the whole particle is supposed to be locked into the solid matrix. The second one is the rigid dipole (RD) model, within which the particle magnetization is supposed to be locked into the crystal lattice. The last approach is very fruitful, well corresponding to the real anisotropic particle of a radius more than $20  \textrm{nm}$ and not very high frequencies ($~ 10^3 - 10^6 \textrm{MHz}$) of external fields \cite{andr2006magnetism}. It is interesting to note that in recent studies of the coupled dynamics \cite{PhysRevB.95.104430}, a large amount of the results was obtained for the RD limit.

Due to small sizes (several tens of nanometers) of the particles, which are contained in a ferrofluid, thermal bath is the first factor defining the response to external fields. Hence, the description of the individual particle motion should be stochastic. To this end, the Langevin and Fokker-Planck formalisms are developed. Since the basic concept was discussed in \cite{0038-5670-17-2-R02, Raikher_1994}, numerous investigations devoted to the RD response to external fields have been conducted. Although, this response was also treated in terms of the complex magnetic susceptibility \cite{PhysRevE.63.011504, Raikher2011, PhysRevE.83.021401, 0953-8984-15-23-313, PhysRevE.82.046310, SotoAquino201546}, it can strongly differ from the prediction of the above mentioned quasi-equilibrium linear model presented in \cite{Rosensweig2002370}. The attention was also paid to the important problem of energy dissipation, which is characterised by the average power loss. Despite the scientific relevance of the results presented, they are miscellaneous and do not permit to estimate from unified positions the influence of all system parameters on the heating process.

Due to the long range character of the inter-particle dipole interaction, the dipole fields are the second factor defining the ferrofluid response to external fields. Hence, we encounter the need to solve a many-body problem. To this end, a few approximate approaches are used, and there is no general solution for this. The most analytical techniques are based on different modifications of the mean field concept \cite{OGRADY1983958, MOROZOV199051, BUYEVICH1992276, PhysRevE.64.041405, PhysRevE.79.021407}, see more results in survey \cite{0034-4885-67-10-R01}. Unfortunately, this approach does not account the nearest neighbor correlations and the possible structure formation discussed in \cite{PhysRevE.53.2509, PhysRevE.57.4535, PhysRevLett.110.148306}. Moreover, the mean field approach is hard to apply when a periodic external field acts. The observed in \cite{OGRADY1983958, MOROZOV199051, BUYEVICH1992276, PhysRevE.64.041405, PhysRevE.79.021407} growth of the quasi-static susceptibility caused by the interaction does not imply the increase in the imaginary part of the complex susceptibility that is confirmed by the measurements of the specific absorption rate \cite{0022-3727-49-29-295001, doi:10.1109/TMAG.2016.2516645, doi:10.1063/1.4974803, Arteagacardona2016636, doi:10.1021/acs.molpharmaceut.5b00866, doi:10.1021/acsnano.7b01762, doi:10.2147/IJN.S141072, doi:10.1166/sam.2017.2948, doi:10.1021/acsnano.7b01762}. In order to calculate correctly the distribution of the local dipole fields, the numerical simulation is demanded.

There are two numerical techniques, namely Monte Carlo (MC) and molecular dynamics (MD) methods \cite{Gould2007Simulation, Rapaport2004MD, Haile1997MD}. Despite easy implementation and low consumption of computational resources, the MC method is useless in the case of time-dependent excitations. Hence, this technique is applied for the investigation of the equilibrium and structure properties \cite{0022-3727-13-7-003, BRADBURY1986745, PhysRevE.59.2424, PhysRevE.71.061203, PhysRevLett.110.148306}. The MD method is based on the integration of the coupled Langevin equations for each RD in the ensemble. This technique has stronger requirements to both the computational equipment and program code, but it is free of restrictions, which are intrinsic to the MC method. Therefore, the MD method is widely used for the description of the properties of ferrofluids. Thus, the ferrofluid structure and initial susceptibility were studied in the works \cite{PhysRevE.66.021405, PhysRevE.75.061405}; work \cite{TANYGIN20124006} reports the numerical results of the self-organization and phase transitions in the aggregated structures; the size distribution impact, dynamical and structural effects were studied in \cite{PhysRevE.88.042315, PhysRevE.92.012306}; at last, the magnetic susceptibility spectra and relaxation properties were treated numerically in \cite{PhysRevE.93.063117, C5SM02679B}. At the same time, the influence of the dipole interaction on the power loss was investigated using the model based on the Landau-Lifshitz equation, where internal damping precession of the magnetic moment is only taken into account \cite{PhysRevB.85.045435, PhysRevB.87.174419, PhysRevB.89.014403, PhysRevB.90.214421}. This approach is valid under some circumstances, but it is used primarily because of the simpler equations of motion. Therefore, the role of the dipole interaction in the energy dissipation of RD placed into a viscous liquid remains unclear.

The aim of the present study is in the following. On the one hand, we need to decompose the impact of the regular component, thermal excitation, and collective behaviour on the response of a ferrofluid to a periodic field. On the other hand, we need to trace the synergy of these factors in the ferrofluid heating in order to get insight about the possibility to control and choose the optimal parameters. The final goal defines the methodology of our analysis and the article structure. Firstly, we exploit the purely dynamical approach and find out the possible analytical solutions of the equations of forced rotational motion for a single RD in a viscous fluid. Secondly, we investigate the statistical properties of a single RD interacting with both external periodic field and thermal bath. Finally, we apply the MD method for the description of the joint motion of the particle ensemble. Within this framework, the Barnes-Hut algorithm \cite{Barnes-Hut-Nature1986} and CUDA technology \cite{Sanders2011CUDA} for facilitation of the local dipole fields calculation are utilized successfully.

\section{DESCRIPTION OF THE MODEL}
\label{Desc}

We consider a ferromagnetic particle of radius $R$, uniform mass density $D$, and magnetization $M$ placed in a fluid of viscosity $\eta$. This particle rotates in the fluid under a magnetic field $\mathbf{H} = \mathbf{H}(t)$, which in a general case consists of the external and dipole parts $\mathbf{H} = \mathbf{H}^{ext} + \mathbf{H}^{dip}$. Further the following assumptions are used. First, the exchange interaction between the atomic magnetic moments is assumed to be dominant. Therefore, the magnitude $|\mathbf{M}| = M$ of the particle magnetization $\mathbf{M}$ can be considered as a constant parameter. Second, the particle radius is assumed to be small enough (less than a few tens of nanometers), and the nonuniform distribution of magnetization is energetically unfavorable. Therefore, the single-domain state, which is characterised by $\mathbf{M} = \mathbf{M}(t)$, takes place. And third, the uniaxial anisotropy magnetic field is assumed to be strong enough, and the particle magnetization is directed along this field. Therefore, $\mathbf{M}$ is locked rigidly into the particle body. In fact, the above assumptions are the ground of the RD model. In the simplest way, the rotational dynamics of RD is described firstly in \cite{doi:10.1063/1.1671698}. If the rotation of the fluid surrounding the particle is neglected, then the particle dynamics is governed by a pair of coupled equations
\begin{subequations}\label{eqs1}
\begin{eqnarray}
    \dot{\mathbf{M}} &=& \boldsymbol{\upomega} \times \mathbf{M},
    \label{eq_a}
    \\[6pt]
    J\dot{\boldsymbol{\upomega}} &=& V\mathbf{M} \times \mathbf{H} - 6\eta V \boldsymbol{\upomega}.
    \label{eq_b}
\end{eqnarray}
\end{subequations}
Here, $\boldsymbol{\upomega} = \boldsymbol{\upomega}(t)$ is the angular velocity of the particle, the overdot denotes the time derivative, $J = (2/5)D VR^{2}$ is the moment of inertia of the particle, $V = (4/3)\pi R^{3}$ is the particle volume (we associate the hydrodynamic volume of the particle with its own volume), and the cross denotes the vector product.

In our study we neglect the hydrodynamical interaction. Such approach is widely used, despite it may be oversimplification in some cases as well as the rigid dipole approximation. However, the results of \cite{doi:10.1021/jp8113978} suggests that the  fluid motion can be not critical for structural properties of ferromagnetic particle ensemble. The other reason to use the simple model lies in our intention to utilize the common and inexpensive facilities for further simulations.

\begin{figure}
\centering
\includegraphics[width=0.6\linewidth]{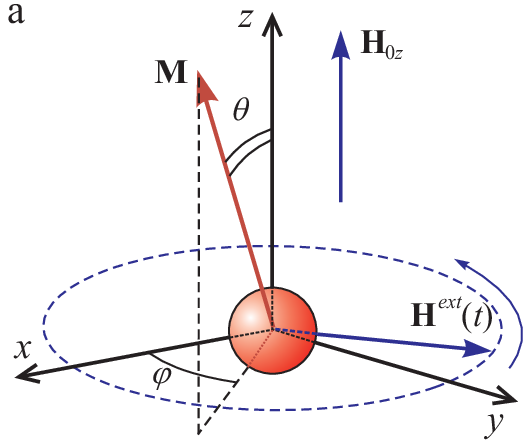}
\includegraphics[width=0.6\linewidth]{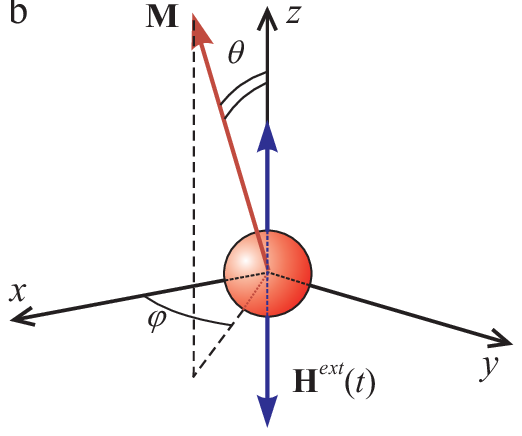}
\caption{(Color online) Schematic representation of our model for the single-particle case. Spherical rigid dipole, its magnetic moment, spherical and Cartesian coordinate systems, and external fields acting on the system are depicted. Plot a: the circularly-polarized field (\ref{eq:ext_fields_cp}) is applied. Plot b: the linearly-polarized field (\ref{eq:ext_fields_cp}) is applied.}
\label{fig:Model}
\end{figure}

Then, we assume that the particle is driven by an external time-periodic field of the following types:
\begin{subequations}\label{eq:ext_fields}
\begin{eqnarray}
    \mathbf{H}^{ext}\!&=&\!H_{m}\big[\mathbf{e}_{x} \cos(\Omega t)\!+\!\mathbf{e}_{y}\varrho \sin(\Omega t)\big]\!+\!\mathbf{e}_{z} H_{0z}, \nonumber \\
    \label{eq:ext_fields_cp}
    \\[6pt]
    \mathbf{H}^{ext}\!&=&\!\mathbf{e}_{z} H_{m}\cos(\Omega t),
    \label{eq:ext_fields_lp}
\end{eqnarray}
\end{subequations}
where $\mathbf{e}_x$, $\mathbf{e}_y$, $\mathbf{e}_z$ are the unit vectors of the Cartesian coordinates, $H_{m}$ is the field amplitude, $\Omega$ is the field frequency, $H_{0z}$ is the static field applied along the $oz$ axis, and $\varrho$ is the factor, which determines the polarization type ($-1 \leq \varrho \leq 1$). The basic concept of the model is sketched in Fig.~\ref{fig:Model}a for a circularly-polarized field (Eq.~(\ref{eq:ext_fields_cp})) and in Fig.~\ref{fig:Model}b for a linearly-polarized field ($\varrho = 1$, Eq.~(\ref{eq:ext_fields_lp})).

\subsection{Equations of motion for Brownian rotation of a single particle}
In the case when the particle size is sufficiently small, the left-hand side of Eq.~(\ref{eq_b}), i.e. the inertia term $J\boldsymbol{\upomega}$, can be neglected for reasonable frequencies. Using this massless approximation and assuming that a random torque $\boldsymbol{\upxi} = \boldsymbol{\upxi}(t)$, which is generated by the thermal bath, is also applied to the particle, one can write
\begin{equation}
    \boldsymbol{\upomega} = \frac{1}{6\eta}\mathbf{M} \times \mathbf{H} + \frac{1} {6\eta V} \boldsymbol{\upxi}.
    \label{omega}
\end{equation}
Substituting the last relation into Eq.~(\ref{eq_a}), we obtain the equation
\begin{equation}
    \dot{\mathbf{M}} = - \frac{1}{6\eta}\mathbf{M} \times (\mathbf{M} \times \mathbf{H}) -
    \frac{1}{6\eta V} \mathbf{M} \times \boldsymbol{\upxi},
    \label{eq:Eq_of_motion_base}
\end{equation}
which describes the stochastic rotation of particles in a viscous fluid. Since the particle magnetization $M$ is constant in time, for further calculations it is reasonable to rewrite Eq.~(\ref{eq:Eq_of_motion_base}) in the spherical coordinates $\mathbf{M} = \mathbf{e}_x M \sin\theta \cos\varphi + \mathbf{e}_y M \sin\theta \sin\varphi + \mathbf{e}_z M \cos\theta$ (see Fig.~\ref{fig:Model})
\begin{subequations}\label{eq:Langevin_angular_base}
\begin{eqnarray}
    \displaystyle \dot{\theta} &=& \frac{1}{\tau_{1}}
    \left(h_x \sin\theta\cos\varphi + h_y \sin\theta\sin\varphi\ + h_z \cos\theta \right) \cot\theta\! -\nonumber
    \\
    \displaystyle &-& \frac{1}{\tau_{1}}\frac{h_z}{\sin\theta} +\!\sqrt{\frac{2}{\tau_{2}}}\left(\zeta_{y}\cos\varphi - \zeta_{x} \sin\varphi\right), \label{eq:Langevin_angular_base_a}
    \\ [10pt]
    \displaystyle \dot{\varphi} &=& \frac{1}{\tau_{1}\sin^{2}\theta}\left(h_{y}\cos\varphi - h_{x} \sin\varphi\right) -\nonumber
    \\
    &-& \sqrt{\frac{2}{\tau_{2}}}\big[(\zeta_{x}\cos \varphi + \zeta_{y}\sin\varphi)\cot\theta - \zeta_{z}\big].
    \label{eq:Langevin_angular_base_b}
\end{eqnarray}
\end{subequations}
Here, $\boldsymbol{\upzeta} = \left(12\eta V k_{\mathrm{B}}T\right)^{-1/2} \boldsymbol{\upxi}$ is the rescaled random torque, $k_\mathrm{B}$ is the Boltzmann constant, $T$ is the absolute temperature, $\tau_{1} = 6\eta/M^{2}$ and $\tau_{2} = 6\eta V/(k_\mathrm{B}T)$ are the characteristic times of the particle rotation induced by the external magnetic field and thermal
torque, respectively, $\mathbf{h} = \mathbf{H}/M$ is the reduced field. The Cartesian components $\zeta_{\nu}$ ($\nu =x,y,z$) of $\boldsymbol{\upzeta}$ are assumed to be independent Gaussian white noises with zero means, $\langle \zeta_{\nu} \rangle =0$, and correlation functions $\langle \zeta_{\nu}(t) \zeta_{\nu}(t') \rangle = \Delta \delta(t-t')$, where $\langle \cdot \rangle$ denotes averaging over all realizations of the Wiener processes $W_{\nu}(t)$ producing noises $\zeta_{\nu}$, $\Delta$ is the dimensionless noise intensity, and $\delta(t)$ is the Dirac delta-function.

An important feature of the Langevin equations (\ref{eq:Langevin_angular_base}) is that noises $\zeta_{\nu}$ are multiplicative ones, i.e. they are multiplied by functions of the angles $\theta$ and $\varphi$. Therefore, the properties of the particle motion can depend on the interpretation of the noises that, in turn, can complicate further processing. To overcome these difficulties, we consider the problem of stochastic rotation of the particle from the statistical point of view. Following \cite{PhysRevE.92.042312}, the Fokker-Planck equation, which corresponds to Eqs.~(\ref{eq:Langevin_angular_base}), is written as
\begin{eqnarray}
    \frac{\partial P}{\partial t} &+& \nonumber \\
    &+& \frac{1}{\tau_{1}} \frac{\partial}{\partial \theta}\bigg[
    \big( h_x \cos\varphi + h_y \sin\varphi \big)\cos\theta  - h_z \sin\theta + \nonumber \\
    &+& \frac{\cot\theta}{\kappa} \bigg]P + \frac{1}{\tau_{1}\sin^{2}\theta}  \frac{\partial}{\partial\varphi}
    \big( h_y \cos\varphi - h_x \sin\varphi \big) P - \nonumber \\
    &-& \frac{1}{\tau_{2}} \frac{\partial^{2}P}{\partial \theta^{2}}-\frac{1}{\tau_{2}} \frac{1}{\sin^{2}\theta} \frac{\partial^{2}P}{\partial \varphi^{2}} = 0,
     \label{eq:FP}
\end{eqnarray}
where $P=P(\theta,\varphi,t)$ is the time-dependent probability density function for the rotational states of the particle, $\kappa=\tau_2/\tau_1 =  M^2 V / k_{\mathrm{B}}T$ is the relationship between the magnetic and thermal energies, which shows the  relative contribution of thermal fluctuations. The latter parameter is the most useful for the analysis of the role of temperature in the particle behavior including energy absorption of the external time-periodic field. By the direct substitution, it is easy to show that the following \textit{effective} Langevin equations
\begin{subequations} \label{eq:Langevin_angular_eff}
\begin{eqnarray}
   \displaystyle \dot{\theta} &=& \frac{1}{\tau_{1}} \big( h_x \cos\varphi + h_y \sin\varphi \big)\cos\theta -
   \frac{1}{\tau_{1}} h_z \sin\theta + \nonumber \\
   \displaystyle&+&\frac{1}{\tau_{2}}\cot \theta + \sqrt{\frac{2}{\tau_{2}}}\,\mu_{1}, \label{eq:Langevin_angular_eff_a}
   \\ [10pt]
   \displaystyle \dot{\varphi} &=& \frac{1}{\tau_{1}}\big( h_y \cos\varphi - h_x \sin\varphi \big)\frac{1}{\sin\theta} + \nonumber \\
   \displaystyle &+& \sqrt{\frac{2}{\tau_{2}}} \frac{1}{\sin\theta}\, \mu_{2}, \label{eq:Langevin_angular_eff_b}
\end{eqnarray}
\end{subequations}
where $\mu_{i} = \mu_{i}(t)$ ($i = 1,2$) are the independent Gaussian white noises with zero means, $\langle \mu_{i}(t) \rangle =0$, and delta correlation functions, $\langle \mu_{i}(t)\mu_{i}(t') \rangle  = \delta(t-t')$, are equal in the statistical sense to the initial equations of motion (\ref{eq:Langevin_angular_base}). At the same time, equations (\ref{eq:Langevin_angular_eff}), in contrast to Eqs.~(\ref{eq:Langevin_angular_base}), do not contain the terms, where the regular functions of angular arguments are multiplied by the noises responsible for fluctuations of the same argument. In other words, only Eq.~(\ref{eq:Langevin_angular_eff_b}) formally contains the multiplicative noise, but here the noise responsible for the fluctuations of the azimuthal angle $\varphi$ is combined with the function of the polar angle $\theta$. Since $\mu_{i}$ are represented by the Gaussian white noises, which are interpreted in the Stratonovich sense, the use of Eqs.~(\ref{eq:Langevin_angular_eff}) instead of Eqs.~(\ref{eq:Langevin_angular_base}) in the numerical simulations is more convenient because of simpler and faster algorithms. It is especially important in a view of the ensemble simulation.

Introducing the dimensionless time $\tilde{t} = t/\tau_{1}$, one can write the system of the \textit{reduced effective} Langevin equations (\ref{eq:Langevin_angular_eff}) as follows
\begin{subequations} \label{eq:Langevin_angular_eff_red}
\begin{eqnarray}
    \displaystyle \frac{d\theta}{d\tilde{t}} &=& \big( h_x \cos\varphi + h_y \sin\varphi \big)\cos\theta -
    h_z \sin\theta + \nonumber\\
    &+& \frac{1}{\kappa}\! \cot\theta + \!\sqrt{\frac{2}{\kappa}} \tilde{\mu}_{1},\label{eq:Langevin_angular_eff_red_a}
    \\ [10pt]
    \displaystyle \frac{d\varphi}{d\tilde{t}} &=& \frac{1}{\sin\theta}\big( h_y \cos\varphi - h_x \sin\varphi \big)-\nonumber\\
    &-& \displaystyle \sqrt{\frac{2}{\kappa}} \frac{1}{\sin\theta}\, \tilde{\mu}_{2},\label{eq:Langevin_angular_eff_red_b}
\end{eqnarray}
\end{subequations}
where $\tilde{\mu}_{i} = \tilde{\mu}_{i}(\tilde{t}) = \sqrt{\tau_{1}}\, \mu_{i}( \tilde{t} \tau_{1})$ ($i=1,2$) are the dimensionless Gaussian white noises with $\langle \tilde{\mu}_{i}(\tilde{t}) \rangle =0$ and $\langle \tilde{\mu}_{i} (\tilde{t}) \tilde{\mu}_{i} (\tilde{t}') \rangle = \delta( \tilde{t} - \tilde{t}')$. It is this system of equations, which is the most convenient for both the analytical and numerical treatment, especially for the large ensemble simulation.

\subsection{Simulation of the interacting particle ensemble}
The interaction is a very important factor that defines the dynamics and properties of a ferrofluid. The problem should to be considered at the following two stages. On the one hand, the resulting dipole fields acting on each particle can influence considerably the rotational dynamics and, in particular, the energy absorbed from the external field. On the other hand, the dipole interaction due to the attractive character induces the cluster structure of the ensemble, that, in turn, impacts the dipole field distribution. The exact analytical description of the particle ensemble driven by time-periodic fields in a viscous liquid is most likely impossible. Therefore, a numerical simulation is demanded. As stated in the introduction, there are two approaches to simulate the ferromagnetic particle ensembles, namely, the MC and MD methods (see \cite{Gould2007Simulation, Rapaport2004MD, Haile1997MD} for details). The last approach is more suitable for high-performance simulation in real time, but it requires large processing power. In this regard, the optimal form of the basic equations, which describe the dynamics of a single particle, plays the key role. And the effective equations (\ref{eq:Langevin_angular_eff_red}) are fit enough for these purposes.

In our approach we expand the model developed above to the case of an ensemble. Thus, here we consider an ensemble of equal spherical uniform ferromagnetic uniaxial particles with the parameters stated above. The rotational motion of the particles is described by the effective stochastic equations similar to Eqs.~(\ref{eq:Langevin_angular_eff_red}), which are complemented by the standard equations of translational motion \cite{PhysRevE.66.021405, Polyakov20131483} written for the massless case as
\begin{subequations} \label{eq:Langevin_angular_eff_red_k}
\begin{eqnarray}
    \displaystyle \frac{d\theta_k}{d\tilde{t}} &=& \big( h_{kx} \cos\varphi_k + h_{ky} \sin\varphi_k \big)\cos\theta_k -
    h_{kz} \sin\theta_k + \nonumber\\
    &+& \frac{1}{\kappa}\! \cot\theta_k + \!\sqrt{\frac{2}{\kappa}} \tilde{\mu}_{k1},\label{eq:Langevin_angular_eff_red_k_a}
    \\ [10pt]
    \displaystyle \frac{d\varphi_k}{d\tilde{t}} &=& \frac{1}{\sin\theta_k}\big( h_{ky} \cos\varphi_k - h_{kx} \sin\varphi_k \big) -\nonumber\\
    &-& \displaystyle \sqrt{\frac{2}{\kappa}}\frac{1}{\sin\theta_k}\, \tilde{\mu}_{k2},\label{eq:Langevin_angular_eff_red_k_b}
    \\ [10pt]
    \displaystyle \frac{d{\boldsymbol{\rho}}_k}{d\tilde{t}} &=& \frac{16 \pi}{9} (\mathbf{f}^{dip}_k + \mathbf{f}^{sr}_k) + \sqrt{\frac{8}{3 \kappa}}\tilde{\mu}_{k3},\label{eq:Langevin_angular_eff_red_k_c}
\end{eqnarray}
\end{subequations}
where $\boldsymbol{\rho}_k$ is the vector defining the reduced (here, the particle radius $R$ is the distance unit) coordinates of the given particle, $k$ is the index number of the given particle in the ensemble, $\mathbf{h}_k = \mathbf{h}_{k}^{dip} + \mathbf{h}^{ext}$ is the resulting dimensionless field acting on the $k$-th particle, which consists of the external uniform part ($\mathbf{h}^{ext} = \mathbf{H}^{ext}/M$) and the resulting reduced dipole field
\begin{equation}
\mathbf{h}_{k}^{dip} = \sum_{j = 1,  j \neq k}^{N}{\frac {4 \pi}{3} \frac {3 \boldsymbol{\rho}_{kj} (\mathbf{u}_j \boldsymbol{\rho}_{kj}) -
\mathbf{u}_j \boldsymbol{\rho}_{kj} ^{\, 2}} {\rho_{kj} ^{\, 5}}},
\label{eq:h_dip_red}
\end{equation}
where $\boldsymbol{\rho}_{kj}$ is the vector joining two particles measured in $R$ units, $\mathbf{u}_j = \mathbf{M}_j/M$ is the reduced magnetic moment of the $j$-th particle; $N$ is the total number of particles.

There are two forces acting on each particle which should be taken into account. Firstly, $\mathbf{f}^{dip}_k$ is the force acting on the $k$-th particle and arising from its dipole interaction with all other particles in the ensemble. Secondly, $\mathbf{f}^{sr}_k$ is the force arising from an anti-aggregation coating, which is widely used in real ferrofluids to prevent the particle aggregation. To represent the explicit form of $\mathbf{f}^{dip}_k$, we applied the standard definition $\mathbf{f}^{dip}_k = \left(\mathbf{u}_k\bigtriangledown_k\right)\mathbf{h}^{dip}_k$. The coating provides the particles repulsion primarily. There are numerous examples in literature, when the force produced by this coating is modeled using the Lennard-Jones potential. This type of potential is the most suitable because it leads to the equilibrium states and prevents the infinite spread of the particles even if the dipole interaction is weak. $\mathbf{f}^{sr}_k = - \bigtriangledown_k W_k$, where $W_k = 4\varepsilon \sum_{j = 1,  j \neq k}^{N}{\left[\left(\sigma/\rho_{kj}\right)^{12} - \left(\sigma/\rho_{kj}\right)^6\right]}$. Here, $\sigma$ is the parameter defining the equilibrium distance between two particles, and $\varepsilon$ is the parameter defining the potential barrier depth. Finally, we rewrite the above forces acting on the particle as
\begin{subequations} \label{eq:Langevin_angular_eff_forces}
\begin{eqnarray}
\mathbf{f}_k^{dip} \!&=&\!  \sum_{j = 1,  j \neq k}^{N} \Biggl[
3 \frac {\boldsymbol{\rho}_{kj} (\mathbf{u}_j \mathbf{u}_k) + \mathbf{u}_k (\mathbf{u}_j \boldsymbol{\rho}_{kj}) +
\mathbf{u}_j (\mathbf{u}_k \boldsymbol{\rho}_{kj})} {\rho_{kj}^{\, 5}} - \nonumber
\\
\!&-&\! 15 \frac {\boldsymbol{\rho}_{kj} (\mathbf{u}_k \boldsymbol{\rho}_{kj})(\mathbf{u}_j \boldsymbol{\rho}_{kj})}
{\rho_{kj}^{\, 7}} \Biggl], \label{eq:Langevin_angular_eff_forces_a}
\\
\mathbf f_k^{sr} \!&=&\!  24 \varepsilon \sum_{j = 1, j \neq k}^{N}
\frac {\boldsymbol{\rho}_{kj}} {\boldsymbol{\rho}_{kj} ^{\, 2}}
\left[ {\left( \frac{\sigma}{\rho_{kj}} \right)} ^{12}
- {\left (\frac{\sigma}{\rho_{kj}} \right) } ^6 \right]. \label{eq:Langevin_angular_eff_forces_b}
\end{eqnarray}\
\end{subequations}
The dipole field calculation is the most computational power consuming part of the numerical algorithm. This is the main factor determining the optimal balance between the computational time ($\mathcal{T}_{sim}$), ensemble size ($N$), and equipment used. The exact direct  calculation of the dipole fields induced by all particles is characterised by the square relationship between time and size ($\mathcal{T}_{sim} \sim N^2$). Instead of the cumbersome exact calculation, two approximations are used. The first approximation is the so-called fast multipole method \cite{GREENGARD280} that provides the performance $\mathcal{T}_{sim} \sim N$, and the second one is the Barnes-Hut algorithm \cite{Barnes-Hut-Nature1986} that provides the performance $\mathcal{T}_{sim} \sim N\log N$. Despite the better facilitation, the fast multipole method does not calculate the neighbour correlations with a good accuracy. Therefore, in our calculation we have utilized  the Barnes-Hut approach. Its main idea consists in the averaging of the fields generated by the particles, which are far enough from the particle under consideration, and, in contrary, the fields generated by the nearest particles are calculated exactly.

Another important feature of our numerical approach is the use of computing capabilities of video cards. This gives an excellent possibility of high-performance calculations on common PC. The video card graphics processing units designed for displaying real time video can be adapted for general-purpose computing. The so-called CUDA technology, which was unveiled by Nvidia company \cite{Sanders2011CUDA}, provides us with the convenient tools for this. Nowadays, many scientific problems can be solved in an inexpensive way and without special facilities like clusters or supercomputers. The collective dynamics of the particle ensembles with the long-range dipole interaction is a suitable problem to demonstrate the abilities of CUDA. The details of the used simulation technique are explained in \cite{Polyakov20131483}.

\subsection{The power loss: definitions and calculation technique}

The dynamics of a particle in a viscous fluid is accompanied by the dissipation of magnetic energy in an external field. We introduce the power loss $Q$, i.e. the magnetic energy dissipation per unit time, in a standard way using the variation of the magnetic energy $\delta W$, which is associated with the magnetic moment increment $\delta \mathbf{M}$ in the external field $\mathbf{H}^{ext}$. Within the assumption that all energy changes are transformed into the irreversible losses, one can write $\delta Q = \mathbf{H}^{ext}\delta \mathbf{M}$. In the simplest noiseless single-particle case, the resulting $Q$ value was obtained by averaging over time
\begin{equation}
    Q = \lim_{\tau \to \infty} \frac{1}{\tau} \int_{0}^{\tau} \mathbf{H}^{ext} \frac{\partial\mathbf{M}}{\partial t}dt.
    \label{eq:def_Q}
\end{equation}
In the reduced form $\widetilde{Q} = Q/(M^2\tau^{-1}_{1})$, which is the quantity of our main interest, the power loss can be written in the form
\begin{equation}
    \widetilde{Q} = \lim_{\widetilde{\tau} \to \infty} \frac{1}{\widetilde{\tau}}
    \int_{0}^{\widetilde{\tau}} \mathbf{h}^{ext}
    \frac{\partial\mathbf{u}}{\partial \widetilde{t}} d\widetilde{t},
    \label{eq:def_Q_red}
\end{equation}
where $\mathbf{u} = \mathbf{M}/M$ is the unit vector of the particle magnetization. It is reasonable to underline here that in the simplest cases of periodic forced motion of $M$, the integration in Eq.~(\ref{eq:def_Q_red}) can be conducted over the reduced field period ($\widetilde{\mathcal{T}} = \mathcal{T}/\tau_1$) only,
\begin{equation}
    \widetilde{Q} = \frac{1}{\widetilde{\mathcal{T}}}
    \int_{0}^{\widetilde{\mathcal{T}}} {\mathbf{h}^{ext}\frac{\partial\mathbf{u}}{\partial\widetilde{t}}} d\widetilde{t}.
    \label{eq:def_Q_red_1}
\end{equation}

In the stochastic case, we need to perform averaging over all the angular states taking into account the probability of each of them. And here, the reduced power loss is calculated as follows
\begin{equation}
    \widetilde{Q} = \lim_{\widetilde{\tau} \to \infty}  \int_{\pi}^{0}d\theta \int_{2\pi}^{0}d\varphi  P(\theta, \varphi, \widetilde{t}) \int_{0}^{\widetilde{\tau}}\mathbf{h}^{ext} \frac{\partial\mathbf{u}} {\partial\widetilde{t}} d\widetilde{t}.
    \label{eq:def_Q_stoch_red}
\end{equation}
Since $\mathbf{u}$ is the stochastic function, the integration in Eq.~(\ref{eq:def_Q_stoch_red}) cannot be carried out in a way, which is common for regular functions. The main difficulty here is in the interpretation of time derivative of $\mathbf{u}$. To avoid this, let us use the well-known approach of integration by parts $\int^{b}_{a} {u(x)v'(x)dx} = \left[U(x)V(x)\right]^{b}_{a} - \int^{b}_{a} {U'(x)V(x)dx}$. Then, we neglect the possible nonlinear effects, such as chaotic \cite{PhysRevB.91.054425} or quasi-periodic \cite{0953-8984-21-39-396002} modes, which occur in the internal magnetic dynamics in the fixed particle. From the numerical solution of equations of motion (\ref{eq:Langevin_angular_base}) for the noiseless case, and equations of motion (\ref{eq:Langevin_angular_eff}) for nonzero temperature, the following conclusions can be done. Firstly, these modes can be generated in a narrow frequency domain, when $\Omega \sim 1/\tau_1$. Secondly, in our case the effects caused by these modes are suppressed by thermal noise on the large time scale. Therefore, we suppose that $\left[\mathbf{u}\mathbf{h}\right]^{0}_{\widetilde{\mathcal{T}}} = 0$, and for the further calculations we use the relationship
\begin{equation}
    \widetilde{Q} = - \lim_{\widetilde{\tau} \to \infty}  \int_{\pi}^{0}d\theta \int_{2\pi}^{0}d\varphi  P(\theta, \varphi, \widetilde{t}) \int_{0}^{\widetilde{\tau}} {\mathbf{u}\frac{\partial\mathbf{h}^{ext}} {\partial\widetilde{t}} d\widetilde{t}}.
    \label{eq:def_Q_stoch_red_1}
\end{equation}

If the probability density $ P(\theta, \varphi, \widetilde{t})$ is the known function of period $\widetilde{\mathcal{T}}$, the integration in Eq.~(\ref{eq:def_Q_stoch_red_1}) can be performed over this period
\begin{equation}
    \widetilde{Q} = - \frac{1} {\widetilde{\mathcal{T}}} \int_{\pi}^{0}d\theta \int_{2\pi}^{0}d\varphi P(\theta, \varphi, \widetilde{t}) \int_{0}^{\widetilde{\mathcal{T}}} \mathbf{u} \frac{\partial\mathbf{h}^{ext}} {\partial\widetilde{t}} d\widetilde{t}.
    \label{eq:def_Q_stoch_red_2}
\end{equation}

For the numerical simulation, the integration in Eq.~(\ref{eq:def_Q_stoch_red_1}) is replaced by the summation, and the corresponding difference scheme is used. Based on the spherical representation of the reduced magnetic moment $\mathbf{u} = \mathbf{e}_x\sin\theta \cos\varphi + \mathbf{e}_y\sin\theta \sin\varphi + \mathbf{e}_z\cos\theta$ and Eq.~(\ref{eq:def_Q_stoch_red_1}), the explicit form of this difference scheme can be written as
\begin{eqnarray}
    \widetilde{Q} &=& \frac{-1} {N_1 N_2} \sum^{N_1 N_2}_{i = 1} \bigg[\sin\theta(\widetilde{t}_i) \cos\varphi(\widetilde{t}_i) \frac {\partial h^{ext}_x(\widetilde{t}_i)} {\partial \widetilde{t}} +  \nonumber\\
    &+& \sin\theta(\widetilde{t}_i) \sin\varphi(\widetilde{t}_i)
    \frac {\partial h^{ext}_y(\widetilde{t}_i)}{\partial \widetilde{t}} + \cos\theta(\widetilde{t}_i)
    \frac {\partial h^{ext}_z(\widetilde{t}_i)}{\partial \widetilde{t}}\bigg]\Delta \widetilde{t}, \nonumber\\
    \label{eq:def_Q_stoch_red_num}
\end{eqnarray}
where $N_1 = \widetilde{\mathcal{T}}/\Delta \widetilde{t}$ is the number of time steps on the external field period, $N_2 = \widetilde{\mathcal{T}}_{sim}/\widetilde{\mathcal{T}}$ is the number of periods, during which the simulation is carried out, $\Delta \widetilde{t}$ is the time increment, which is constant in the simulation.

Finally, in the case of the interacting ensemble composed of $N$ particles and simulated on the basis of Eqs.~(\ref{eq:Langevin_angular_eff_red_k}), we need to conduct the additional averaging over all particles in the ensemble. We update the technique used in Eq.~(\ref{eq:def_Q_stoch_red_num}), and the resulting expression applied for calculation of the power loss in the interacting ensemble has the following form:
\begin{eqnarray}
    \displaystyle \widetilde{Q}\!&=&\!\frac{-1} {N_1 N_2 N} \sum^{N}_{k = 1} \sum^{N_1 N_2}_{i = 1}  \bigg[\sin\theta_k(\widetilde{t}_i) \cos\varphi_k(\widetilde{t}_i) \frac {\partial h^{ext}_{kx}(\widetilde{t}_i)} {\partial \widetilde{t}}\!+  \nonumber\\
    \!&+&\!\displaystyle \sin\theta_k(\widetilde{t}_i) \sin\varphi_k(\widetilde{t}_i)
    \frac {\partial h^{ext}_{ky}(\widetilde{t}_i)}{\partial \widetilde{t}}\!+\!\cos\theta(\widetilde{t}_i)
    \frac {\partial h^{ext}_{kz}(\widetilde{t}_i)}{\partial \widetilde{t}}\bigg]\Delta \widetilde{t}. \nonumber\\
    \label{eq:def_Q_stoch_red_num_1}
\end{eqnarray}
We suppose that the decrease in the external field energy is simultaneously compensated from the external field source. Moreover, we do not take into account the energy increments arising from the changes of the dipole field $\mathbf{h}_{k}^{dip}$. The increase in the energy of the given particle with changing dipole field is accompanied by the same decrease in the energy of other particles, which are the sources of this dipole field. In other words, the dipole field can transfer energy from one particle to another, but cannot produce the additional power loss.

The details of the numerical calculation in the present paper are the following. To simulate the single-particle stochastic dynamics, the system of equations (\ref{eq:Langevin_angular_eff_red}) was solved by the second-order Runge-Kutta method with the time quantification step of $\Delta \widetilde{t} = 0.005 \widetilde{\mathcal{T}} $ in the range of $N_2 = 1000$ reduced field periods for each point of the plot. To simulate the behaviour of the interacting ensemble, the system of equations (\ref{eq:Langevin_angular_eff_red_k}) was solved in the same way with the time quantification step of $\Delta \widetilde{t} = 0.005 \widetilde{\mathcal{T}} $ in the range of $N_2 = 1000$ reduced field periods for each point of the plot for $N = 4096$ number of particles. The values of other system parameters are specified below. The video-cards Nvidia GeForce 450 GTS and Nvidia GeForce 650 GTS Ti were used for our simulation. The program code was realized using C++ language and Eclipse development environment.

\section{RESULTS AND DISCUSSION}
\label{Res}
\subsection{Single-particle noiseless case}
In the simplest case of a diluted ferrofluid, when each particle is far enough from its neighbours and when the thermal energy is much smaller than the magnetic one ($\kappa \gg 1$), the equations of motion (\ref{eq:Langevin_angular_base}) can be solved exactly  in some specific cases. These results have a direct practical value. Firstly, they can be applied to the power loss calculation under the mentioned circumstances. Secondly, they establish the limit values for the stochastic and interacting cases. The last, but not least, is the methodological relevance of the results obtained. The presence of the analytical solutions in the simplest case lets us verify the processing methods of more complicated cases. Despite some results for the regular dynamics of a rigid dipole in a viscous fluid were obtained earlier (see Refs. \cite{Raikher2011, PhysRevE.83.021401, Lyutyy201887}), it is reasonable to systematize and generalize all of them here.
\subsubsection{Motion in the linearly-polarized field}
Let us consider the linearly-polarized field action firstly. Since the field given by Eq.~(\ref{eq:ext_fields_lp}) oscillates along the $oz$-axis only, the azimuthal angle $\varphi$ will remain constant and $\dot{\varphi}=0$. Therefore, Eq.~(\ref{eq:Langevin_angular_base_b}) can be neglected. Substituting Eq.~(\ref{eq:ext_fields_lp}) into Eq.~(\ref{eq:Langevin_angular_base_a}) within the assumption $\kappa \rightarrow \infty$, we obtain a rather simple differential equation
\begin{equation}
\displaystyle \frac{d\theta}{d\widetilde{t}} = - h_{m}\sin\theta \cos(\widetilde{\Omega}\tilde{t}).
\label{eq:dinam_lp_base_eq}
\end{equation}
Eq.~(\ref{eq:dinam_lp_base_eq}) can be integrated directly, and the expression describing the particle spherical motion is written as
\begin{equation}
    \tan (\theta/2) = \tan (\theta_0/2)\exp\left[-\dfrac{h_{m}}{\widetilde{\Omega}}\sin(\widetilde{\Omega}\tilde{t})\right],\\
    \label{eq:dinam_lp_solution}
\end{equation}
where $\theta_0$ is the initial polar angle of vector $\mathbf{u}$, $h_{m} = H_{m}/M$ is the reduced field amplitude. As obvious from Eq.~(\ref{eq:dinam_lp_solution}), the scale of particle oscillations is very sensitive to the ratio $h_{m}/\widetilde{\Omega}$. When $h_{m}/\widetilde{\Omega} \gg 1$, the particle reorientation along the external field is performed fast enough, and the particle is practically immobilized during the most part of the field period. In contrary, when $h_{m}/\widetilde{\Omega} \ll 1$, only small oscillations are performed around the initial position, which is defined by $\theta_0$.

Using (\ref{eq:def_Q_red_1}), the average value of the power loss under the linearly-polarized field action can be written in the form
\begin{equation}
    \widetilde{Q} = \dfrac{\widetilde{\Omega}^{2}h_{m}}{2\pi} \int_{0}^{\widetilde{\mathcal{T}}} d\widetilde{t} \tanh \left[ \dfrac{h_{m}}{\widetilde{\Omega}}\sin(\widetilde{\Omega} \tilde{t}) - \dfrac{x_0}{2}\right] \sin (\widetilde{\Omega} \tilde{t}),
    \label{eq:Q_lin}
\end{equation}
where $x_0 = \ln \left(\tan^2 (\theta_0/2\right))$ is the constant defined by the initial state of $\mathbf{u}$. Despite the exact integration is impossible here, the value of the power loss given by Eq.~(\ref{eq:Q_lin}) can be evaluated approximately in the limits of low and high frequencies. Thus, when $h_{m}/\widetilde{\Omega} \gg 1$, the power loss demonstrates a linear behaviour with respect to the field frequency and amplitude, $\widetilde{Q}|_{h_{m}/\widetilde{\Omega}\rightarrow \infty} \rightarrow h_{m}\widetilde{\Omega}/\pi$. Another feature of this asymptotics is an independence on the initial position of $\mathbf{u}$ that is explained as follows. During the field period, the particle has enough time to perform two reorientations along with the external field of the type Eq.~(\ref{eq:ext_fields_lp}) from any initial position. Therefore, the stationary mode is not sensitive to $\theta_0$. The high-frequency asymptotics ($h_{m}/\widetilde{\Omega} \ll 1$) is characterized by an independence on the field frequency, $\widetilde{Q}|_{h_{m}/\widetilde{\Omega}\rightarrow 0} \rightarrow 0.5 h_{m}^2 \cosh^{-2}x_0$. The independency on the frequency explained as follows. When the condition $h_{m}/\widetilde{\Omega}\rightarrow 0$ holds, the oscillations is small. The power loss is defined by the the oscillation amplitude and frequency. The latter is equal to the external field frequency, while the former is inversely to it. And the ﬁeld frequency increase leads to the proportional decrease in the $\mathbf{u}$ oscillation amplitude that compensates the power loss increase.

\subsubsection{Precession in the circularly-polarized field}
Then we consider the circularly-polarized field action. Using again the condition $\kappa \rightarrow \infty$ for the noiseless assumption together with the representation of the circularly-polarized field Eq.~(\ref{eq:ext_fields_cp}), we transform Eqs.~(\ref{eq:Langevin_angular_base}) into the set of differential equations
\begin{equation}
    \begin{array}{ll}
    \displaystyle \frac{d\theta}{d\widetilde{t}} = h_{m}\cos\theta \cos(\widetilde{\Omega}\tilde{t}-\varphi) - h_{0z} \sin\theta,
    \\ [10pt]
    \displaystyle \frac{d\varphi}{d\widetilde{t}} = h_{m}\frac{\sin(\widetilde{\Omega}\tilde{t}-\varphi)}{\sin\theta},
    \end{array}
    \label{eq:dinam_cp_base_eq}
\end{equation}
where $h_{0z} = H_{0z}/M$. One of the possible modes of motion is the precession of the vector $\mathbf{u}$ along with the external field $\mathbf{h}^{ext}$. In this case, the solution of Eqs.~(\ref{eq:dinam_cp_base_eq}) is given in the form  of $\varphi = \varrho \widetilde{\Omega}\tilde{t} -\Phi$ and $\theta = \Theta$. The direct substitution of the las formulas into Eqs.~(\ref{eq:dinam_cp_base_eq}) permits us to write the set of algebraical equations
\begin{equation}
    \begin{array}{ll}
    h_{m}\cos\Theta \cos\Phi - h_{0z} \sin\Theta = 0,
    \\ [10pt]
    \widetilde{\Omega} \sin \Theta  - h_{m}\sin\Phi = 0,
    \end{array}
   \label{eq:dinam_cp_solution}
\end{equation}
the solution of which exhaustively describes the precessional mode. It is important to note that this mode remains stable, when $h_{0z} \neq 0$ or when $h_{0z} = 0$ and $h_{m}\widetilde{\Omega} < 1$. The straightforward calculations using Eqs.~(\ref{eq:dinam_cp_solution}) and Eq.~(\ref{eq:def_Q_red_1}) yield the following value of the power loss:
\begin{equation}
    \widetilde{Q} = \widetilde{\Omega}^{2} \sin^{2}\Theta.
    \label{eq:Q_cp}
\end{equation}
In the case of the static field absence ($h_{0z} = 0$), vector $\mathbf{u}$ rotates in the $xoy$ plane and Eq.~(\ref{eq:Q_cp}) is reduced into the expression $\widetilde{Q} = \widetilde{\Omega}^{2}$. It is notably that the power loss here does not depend on the field amplitude $h_{m}$.

\subsubsection{Small oscillations around the initial position of $\mathbf{u}$}
\begin{figure}
\centering
\includegraphics[width=0.6\linewidth]{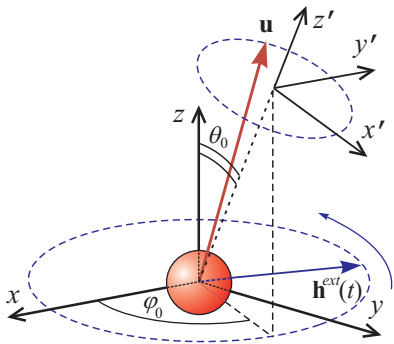}
\caption{(Color online) Schematic representation of our model in the case of small oscillations of the single particle. The further analysis is performed in the primed coordinate system rotated by angles $\varphi_0$, $\theta_0$, which define the initial position of the particle. As an example, the external field is supposed to be circularly-polarized, but the consideration remains valid for other polarizations}
\label{fig:Model_new_coord}
\end{figure}

Finally, we consider the limit case when the vector $\mathbf{u}$ performes rotational oscillations in a small vicinity around its initial position defined by the angles $\theta_0$ and $ \varphi_0$ (see Fig.~\ref{fig:Model_new_coord}). This situation takes place for a small enough ratio of the field amplitude and frequency ($h_{m}/\widetilde{\Omega} \ll 1$). Then, we suppose that the external field is defined as Eq.~(\ref{eq:ext_fields_cp}), but in addition we assume that $h_{0z} = 0$ and $-1 < \varrho < 1$ that includes the linear, elliptical, and circular polarization of $\mathbf{h}^{ext}$. The solution of the basic equations (\ref{eq:Eq_of_motion_base}) in the noiseless limit can be found in the linear approximation. The linearization procedure used here is similar to that reported in \cite{PhysRevB.91.054425} and consists in the following. We introduce the primed coordinate system $x' y' z'$, which is rotated by the angles $\theta_0$ and $\varphi_0$ with respect to the laboratory system $xyz$. In this new coordinate system, the vector $\mathbf{u}$ can be written in the linear approximation as
\begin{equation}
\mathbf{u}=\mathbf{e}_{x'}u_{x'}+\mathbf{e}_{y'}u_{y'}+\mathbf{e}_{z'},
\label{eq:m_lin_gen_sol}
\end{equation}
where $\mathbf{e}_{x'}, \mathbf{e}_{y'}, \mathbf{e}_{z'}$ are the unit vectors of the coordinate system $x' y' z'$.
The external field $\mathbf{h}^{ext}$ in the primed coordinate system $x' y' z'$ can be represented using the known rotation matrix as follows
\begin{equation}
\begin{array}{lcl}
    {\mathbf{h}^{ext}}' = \mathbf{C}\cdot
        \left(
        \begin{array}{c}
            h_{m} \cos(\widetilde{\Omega}\tilde{t}) \\
            \varrho h_{m} \sin(\widetilde{\Omega}\tilde{t})\\
            0 \\
        \end{array}
    \right),
    \label{eq:h_C}
\end{array}
\end{equation}
\begin{equation}
    {\mathbf{C}} =
        \left(
        \begin{array}{lcr}
            \cos\theta_0 \cos\varphi_0 & \cos\theta_0 \sin\varphi_0 & - \sin\theta_0 \\
            -\sin\varphi_0  & \cos\varphi_0 & 0   \\
            \sin\theta_0 \cos\varphi_0 &  \sin\theta_0 \sin\varphi_0 & \cos\theta_0 \\
        \end{array}
        \right),
        \label{eq:C}
\end{equation}
Substituting Eq.~(\ref{eq:h_C}) into Eq.~(\ref{eq:Eq_of_motion_base}) within the assumption that $ u_{x'}, u_{y'} \sim h_{m}$ and neglecting all the terms, which contain $h_{m}$ in any power greater than one, we derive the linearized system of equations for $\mathbf{u}$ in the form
\begin{equation}
\begin{array}{l}
    \dfrac{d u_{x'}}{d\tilde{t}}= \cos\theta_0 \cos\varphi_0\ cos(\widetilde{\Omega}\tilde{t}) + \varrho\!\cos\theta_0 \sin\varphi_0 \sin (\widetilde{\Omega}\tilde{t}),\\
    [2pt]
    \dfrac{d u_{y'}}{d\tilde{t}}= - \sin\varphi_0 \cos(\widetilde{\Omega}\tilde{t}) + \varrho\!\cos\varphi_0 \sin(\widetilde{\Omega}\tilde{t}).\\
\end{array}
 \label{eq:Eq_of_motion_base_lin}
\end{equation}
Then, we use the standard trigonometric representation of the solution of Eqs.~(\ref{eq:Eq_of_motion_base_lin})
\begin{equation}
\begin{array}{lcl}
    u_{x'}= a\cos(\widetilde{\Omega}\tilde{t}) + b\sin(\widetilde{\Omega}\tilde{t}), \\
    [2pt]
    u_{y'}= c\cos(\widetilde{\Omega}\tilde{t}) + d\sin(\widetilde{\Omega}\tilde{t}). \\
\end{array}
\label{eq:Eq_of_motion_base_lin_sol}
\end{equation}
After direct substitution of Eqs.~(\ref{eq:Eq_of_motion_base_lin_sol}) into Eqs.~(\ref{eq:Eq_of_motion_base_lin}), one can easily obtain the unknown constants
\begin{equation}
\begin{array}{lcl}
    a \!\!&=&\!\! h_{m}\widetilde{\Omega}^{-1} \sin \varphi_{0}, \\
    [2pt]
    b \!\!&=&\!\! h_{m}\widetilde{\Omega}^{-1} \cos \theta_{0} \cos \varphi_{0}, \\
    [2pt]
    c \!\!&=&\!\! - h_{m}\widetilde{\Omega}^{-1} \cos \varphi_{0}, \\
    [2pt]
    d \!\!&=&\!\! h_{m}\widetilde{\Omega}^{-1} \cos \theta_{0} \sin \varphi_{0}. \\
\end{array}
\label{eq:Eq_of_motion_base_lin_sol_coef}
\end{equation}
And at last, we integrate directly Eq.~(\ref{eq:def_Q_red_1}) substituting Eqs.~(\ref{eq:Eq_of_motion_base_lin_sol}) and Eqs.~(\ref{eq:Eq_of_motion_base_lin_sol_coef}) and obtain the short expression for the desired power loss
\begin{equation}
    \widetilde{Q} = 0.5 h_{m}^2 D,
    \label{eq:Q_small_osc}
\end{equation}
where
\begin{equation}
    D = \cos^2 \theta_0(\cos^2 \varphi_0 + \varrho^2\sin^2 \varphi_0) + \varrho^2\cos^2 \varphi_0 + \sin^2 \varphi_0.
    \label{eq:D}
\end{equation}
It is natural that Eq.~(\ref{eq:Q_lin}) coincides with the high-frequency asymptotic of Eq.~(\ref{eq:Q_small_osc}) up to a constant.

As an intermediate conclusion we want to emphasize that the results obtained in the dynamical approximation set the limit values for the power loss derived in other approximations. But, as it will be shown below, these estimations for high frequencies can have a practical meaning. Then, we need to underline that the expressions found in the dynamical approximation depend strongly on the initial conditions for all cases, excluding the uniform precession mode under the circularly-polarized field action. And the main feature is that the frequency behaviour of the power loss is not similar to the analogous obtained for the case of magnetic dynamics inside the uniaxial particle, which is fixed rigidly in the solid matrix \cite{PhysRevB.91.054425}. Firstly, for low frequencies the dependencies $\widetilde{Q}(\widetilde{\Omega})$ are quite different for various types of field polarization. When the circularly-polarized field is applied, then $\widetilde{Q} = \widetilde{\Omega}^2$, while $\widetilde{Q} \sim \widetilde{\Omega} h_{m}$ if the linearly-polarized field is applied. Secondly, for high frequencies $\widetilde{Q}(\widetilde{\Omega})$ demonstrates the same saturated character for all polarization types and tends to non-zero constants, but its values for the circularly-polarized field are, at least, two times larger than for the linearly-polarized one depending on the initial position of $\mathbf{u}$.

\subsection{A single particle in the thermal bath}

Obviously that thermal fluctuations blur the rotational trajectories of the particle and suppress its response to the external field. This is a reason of the power loss decay with temperature. And the rate of this decay is of great interest from both the theoretical and practical viewpoints. To this end, the direct integration of the equations of motion, which are stochastic here, is not suitable. Therefore, the analytical estimations are conducted statistically using the probability density function and Fokker-Planck formalism, see Eq.~(\ref{eq:FP}). Because of the difficulties in the exact integration of Eq.~(\ref{eq:FP}) in the case of the time-periodic field action, its solution is often sought in different approximations, such as the effective field approximation \cite{Raikher_1994}, where the form of the distribution corresponds to that of the distribution in a static field or to the steady-state solution in the linear approximation in $\kappa \widetilde{\Omega}$ \cite{PhysRevE.92.042312}. It is remarkable that in the case of the linearly-polarized field Eq.~(\ref{eq:ext_fields_lp}) action, the Fokker-Planck equation (\ref{eq:FP}) can be found exactly in the form of series \cite{0953-8984-15-23-313}. Here we summarize all the results in the context of the power loss problem and confirm them numerically based on Eqs.~(\ref{eq:Langevin_angular_eff_red}).

\subsubsection{Random motion in the linearly-polarized field}
Firstly let us consider the case, when the external field oscillates along the $oz$-axis, see Eq.~(\ref{eq:ext_fields_lp}). Due to the symmetry reasons, we suppose that the probability density function $P$ depends on the polar angle $\theta$ only. Then, following \cite{0953-8984-15-23-313}, we present $P$ in the form of $P = P(\theta, \tilde{t}) = \sin\theta f(\tilde{t})$ that, in turn, allows to transform the Fokker-Planck equation (\ref{eq:FP}) into
\begin{eqnarray}
  \dfrac {df}{d\tilde{t}} &=& \nonumber \\
   &=& \dfrac {1}{\kappa} \bigg[\dfrac {1}{\sin\theta} \dfrac {\partial}{\partial\theta}\bigg(\sin\theta\dfrac {\partial f}{\partial\theta} + f\kappa h_{m} \sin^2\theta\cos(\widetilde{\Omega} \tilde{t})\bigg)\bigg].\nonumber \\
   \label{eq:f_lp_eq}
\end{eqnarray}
To simplify the further calculations, we use the designation $\cos\theta = x$. Taking into account the latter, we write finally
\begin{eqnarray}
  \dfrac {df}{d\tilde{t}} \!&=&\! \nonumber \\
   \!&=&\! \dfrac {1}{\kappa} \dfrac {\partial}{\partial x}\bigg[(1-x^2)\dfrac {\partial f}{\partial x} - \kappa h_{m} \cot(\widetilde{\Omega} \tilde{t})(1-x^2)f \bigg].\nonumber \\
   \label{eq:f_lp_eq_1}
\end{eqnarray}
We underline that equation (\ref{eq:f_lp_eq_1}) coincides entirely with the corresponding expression in \cite{0953-8984-15-23-313}. Its solution was obtained by expansion in Legendre polynomials and harmonics
\begin{eqnarray}
   f \!&=&\! \nonumber \\
   \!&=&\! 0.5\!+\!\sum^{\infty}_{\ell=0} {\bigg[ \sum^{\infty}_{n=0}{A_{\ell n}\cos(n\widetilde{\Omega} \tilde{t})}\!+\! \sum^{\infty}_{n=0}{B_{\ell n}\sin(n\widetilde{\Omega} \tilde{t})} \bigg]}\mathcal{P}_{\ell}(x),\nonumber \\
   \label{eq:f_sol}
\end{eqnarray}
where $\mathcal{P}_{\ell}(x)$ are the Legendre polynomials, $n$, $\ell$ are the whole numbers. Direct substitution of Eq.~(\ref{eq:f_sol}) into Eq.~(\ref{eq:f_lp_eq_1}) lets us to derive the algebraic set of equations, which yields the unknown coefficients $A_{\ell n}$, $B_{\ell n}$
\begin{eqnarray}
   -n\widetilde{\Omega} A_{\ell n} \!&=&\! \nonumber \\
   \!&=&\!\dfrac{1}{\kappa}\bigg[\!-\!\ell(\ell + 1) B_{\ell n}\!+\!\dfrac{\kappa h_{m}}{2}\bigg( \dfrac{\ell(\ell+1)}{2\ell-1}\big(B_{\ell-1,n-1}\!+\! \nonumber \\ \!&+&\!B_{\ell-1,n+1}\big)-\dfrac{\ell(\ell+1)}{2l+3}\big(B_{\ell+1,n-1}\!+\!B_{\ell+1,n+1}\big)\bigg)\bigg],
   \nonumber \\
   \label{eq:A_ln}
\end{eqnarray}
\begin{eqnarray}
   n\widetilde{\Omega} B_{\ell n} \!&=&\! \nonumber \\
   \!&=&\! \dfrac{1}{\kappa}\bigg[\!-\!\ell(\ell + 1) A_{\ell n}\!+\!\nonumber \\
   \!&+&\! \dfrac{\kappa h_{m}}{2}\bigg( \dfrac{\ell(\ell+1)}{2\ell-1}\big((1\!+\!\delta_{n \ell})A_{\ell-1,n-1} + A_{\ell-1,n+1}\big)\!- \nonumber \\
   \!&-&\! \dfrac{\ell(\ell+1\!)}{2\ell+3}\big((1+\delta_{n \ell})A_{\ell+1,n-1}\!+\!A_{\ell+1,n+1}\big)\bigg)\bigg],
   \nonumber \\
   \label{eq:B_ln}
\end{eqnarray}
when  $n \geq 1$ or $\ell \geq 1$. Other cases are defined as follows
\begin{equation}
   A_{\ell 0} = \dfrac{\kappa h_{m}}{2}\bigg[ \dfrac{1}{2\ell-1}A_{\ell - 1, 1} - \dfrac{1}{2\ell+3}A_{\ell + 1, 1}\bigg]
      \label{eq:A_l0_ln}
\end{equation}
for $\ell \geq 1$, $A_{00}  = 0.5$, $A_{0n}  = 0$, $B_{\ell 0}  = 0$, $B_{0n}  = 0$, and, finally, $A_{\ell n} = 0$, $B_{\ell n} = 0$, when $\ell + n$ is odd.

\begin{figure}
\centering
\includegraphics[width=1.0\linewidth]{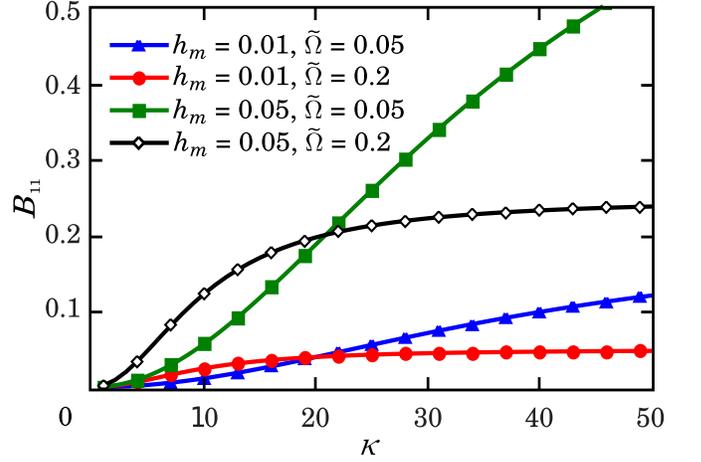}
\caption{(Color online) The numerically obtained dependencies of the coefficient $B_{11}$ in Eqs.~(\ref{eq:B_ln}) on the parameter $\kappa$, which represents the relationship between the magnetic and thermal energies. The saturated character and complex frequency behaviour are in focus}
\label{fig:onepart_res_lp_B_11}
\end{figure}

The power loss $\widetilde{\Omega}$ in the case of the linearly-polarized field action was obtained by the direct substitution of Eq.~(\ref{eq:f_sol}) (taking into account that $P = P(\theta, \tilde{t}) = \sin\theta f(\tilde{t})$) into Eq.~(\ref{eq:def_Q_stoch_red_2}). After performing all integration procedures, the ultimate expression has a rather simple form
\begin{equation}
   \widetilde{Q} = \dfrac{1}{3} h_{m} \widetilde{\Omega} B_{11},
   \label{eq:Q_stoch_lp}
\end{equation}
which correlates with the noiseless analogue, see the analysis of Eq.~(\ref{eq:Q_lin}). The noise influence is hidden in the parameter $B_{11}$, which can be analyzed only in a numerical way. As seen from Fig.~\ref{fig:onepart_res_lp_B_11}, the dependencies $B_{11} (\kappa)$, which were obtained through the numerical treatment of Eqs.~(\ref{eq:A_ln})-(\ref{eq:A_l0_ln}), are saturated for large frequencies. Then, for small $\kappa$, the values of $B_{11}$ increase rapidly at large frequencies, while for large $\kappa$ these trends are changed to the opposite. This fact suggests about the advantage of the high-frequency fields use, when the temperature is high. Finally, $B_{11} (\kappa)$ grows with the field amplitude, especially for big $\kappa$. Therefore, the sensitivity of $B_{11}$ to $\kappa$ increases with $h_{m}$ and $\widetilde{\Omega}$.

\begin{figure}
\centering
\includegraphics[width=1.0\linewidth]{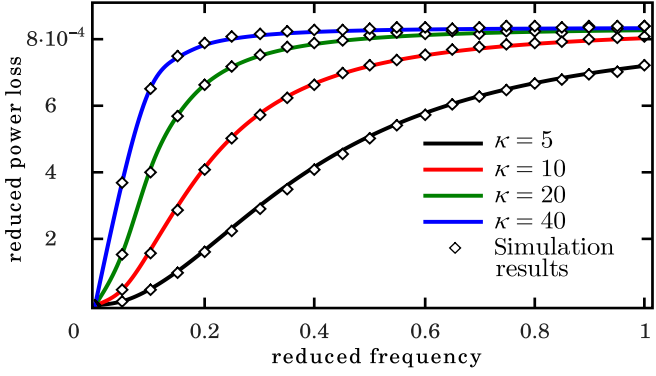}
\caption{(Color online) The frequency dependencies of the power loss for different values of the relationship between the magnetic and thermal energies while the linearly-polarized field is applied. The value of the field amplitude is chosen as $h_{m} = 0.05$. The numerical and analytical predictions are in a good agreement. The saturated character of the behaviour and the correspondence with the noise-free results are highlighted}
\label{fig:onepart_res_lp_num}
\end{figure}

To confirm the analytical findings, the set of simulation runs has been performed on the basis of the numerical integration of Eqs.~(\ref{eq:Langevin_angular_eff_red}). As we can see from Fig.~\ref{fig:onepart_res_lp_num}, the analytical and numerical results are in a good agreement in a wide range of parameters. Then, the dependencies are qualitatively similar to the noiseless limit, and the difference decreases with $\kappa$. At last, this difference is more pronounced for small frequencies and vanishes for large ones.

\subsubsection{Random precession in the circularly-polarized field}
In the case of the circularly-polarized field action, the approximate solution of the Fokker-Planck equation (\ref{eq:FP}) is grounded on the synchronous (in the average sense) rotation of $\mathbf{u}$ along with $\mathbf{h}^{ext}$ \cite{PhysRevE.92.042312}. The transition from the azimuthal angle $\varphi$ to the lag angle $\psi = \varphi - \widetilde{\Omega} \tilde{t}$ permits us to represent the steady-state solution $P_{\mathrm{st}}$ of the Fokker-Planck equation (\ref{eq:FP}) as a function of two variables $P_{\mathrm{st}} =P_{\mathrm{st}} (\theta, \psi)$. Since $\partial P_{\mathrm{st}}/\partial \tilde{t} = \widetilde{\Omega} \partial P_{\mathrm{st}}/\partial \psi$, we can find the equation for the steady-state probability density $P_{\mathrm{st}}$ directly from Eq.~(\ref{eq:FP})
\begin{eqnarray}
    \widetilde{\Omega}\dfrac{\partial P_{\mathrm{st}}}{\partial\psi} &+& \nonumber \\
    &+& \frac{\partial}{\partial \theta}\bigg[h_{m}\cos\psi\cos\theta - h_z \sin\theta + \frac{\cot\theta}{\kappa}\bigg]P_{\mathrm{st}} + \nonumber \\
    &+& \sin^{2}\theta \frac{\partial}{\partial\psi}h_{m}\sin\psi P_{\mathrm{st}} - \frac{1}{\kappa}\frac{\partial^{2}P_{\mathrm{st}}}{\partial \theta^{2}} - \nonumber \\
    &-& \frac{1}{\kappa}\frac{1}{\sin^{2}\theta} \frac{\partial^{2}P_{\mathrm{st}}}{\partial\psi^{2}} = 0.
    \label{eq:FPst}
\end{eqnarray}
Following \cite{PhysRevE.92.042312} and assuming that $\kappa \widetilde{\Omega} \ll 1$, the steady-state probability density $P_{\mathrm{st}}$ is represented in the linear approximation in $\kappa \widetilde{\Omega}$
\begin{equation}
    P_{\mathrm{st}} = (1 + \kappa \widetilde{\Omega} F)P_{0},
    \label{eq:P_st}
\end{equation}
where
\begin{eqnarray}
    P_{0} \!&=&\! \frac{1}{Z} \sin\theta \exp \left[\kappa h_{m}(\sin\theta \cos\psi\!-\!h_{z}\cos\theta)\right],\nonumber \\
    Z \!&=&\! \int_{0}^{\pi}\!d\Theta\!\int_{0}^{2\pi}\!d\psi \sin \theta \exp\left[\kappa h_{m}(\sin\theta \cos\psi\!-\!h_{z}\cos\theta)\right].\nonumber \\
    \label{eq:P_0}
\end{eqnarray}
In what follows, we restrict ourselves to the case, when $h_{z} = 0$ and $\kappa h_{m} \ll 1$. Then, using Eqs.~(\ref{eq:P_0}) and (\ref{eq:FPst}), it is easy to show that in the first-order approximation in $\kappa h_{m}$ the unknown function $F$ is governed by the equation
\begin{equation}
    \frac{1}{\sin\theta} \frac{\partial} {\partial\theta}\bigg(\!\sin\theta\frac{\partial F}{\partial\theta}\bigg)
    + \frac{1}{\sin^{2}\theta} \frac{\partial^{2} F}{\partial \psi^{2}} = - \kappa h_{m}\sin\theta\sin\psi.
    \label{eq:F_eq}
\end{equation}
The solution of this equation is simple enough
\begin{equation}
    F = 0.5 \kappa h_{m} \sin\theta \sin\psi.
    \label{eq:F_sol}
\end{equation}
Taking into account that up to quadratic order in $\kappa h_{m}$, the normalization constant in Eqs.~(\ref{eq:P_0}) can be written as $Z = 4\pi (1 + \kappa^{2}h^{2}/6)$ and
\begin{eqnarray}
    P_{0} &=& \frac{\sin\theta}{4\pi}\! \Big[1 + \kappa h_{m}\sin\theta \cos\psi - \nonumber \\
    &-& \frac{\kappa^{2}h_{m}^{2}}{6}\left(1 - 3\sin^{2}\theta \cos^{2}\psi\right) \Big]\!,
    \label{eq:P_0_1}
\end{eqnarray}
from Eq.~(\ref{eq:P_st}) one immediately gets
\begin{eqnarray}
   P_{\mathrm{st}} &=& \nonumber \\
   \!&=&\! \dfrac{\sin\theta}{4 \pi} \bigg[1\!+\!\kappa h_{m} \sin\theta \cos\psi\!-\!\dfrac{\kappa^2 h_{m}^2}{6}\big(1\!-\!\nonumber \\
   \!&-&\!3 \sin^2\theta \cos^2\psi\big)\bigg]\!+\!\dfrac{1}{8 \pi} \kappa^2 h_{m} \widetilde{\Omega} \sin^2\theta \sin\psi. \nonumber \\
   \label{eq:P_st_fin}
\end{eqnarray}

Finally, the power loss $\widetilde{\Omega}$ in the case of the circularly-polarized field action is also obtained by the direct substitution of Eq.~(\ref{eq:P_st_fin}) into Eq.~(\ref{eq:def_Q_stoch_red_2}). After all, we derive the following formula
\begin{equation}
   \widetilde{Q} = \dfrac{1}{6} h_{m}^2\widetilde{\Omega}^2 \kappa^2.
   \label{eq:Q_stoch_cp}
\end{equation}
The main feature of the last expression is in the quadratic dependence on $h_{m}$, which is not typical for the noiseless case, see Eq.~(\ref{eq:Q_cp}). Then, attention deserves the quadratic dependence on $\kappa$ that denotes the power loss decay on temperature as $T^2$. Here, we underline that Eq.~(\ref{eq:Q_stoch_cp}) is not applicable for small intensities of thermal noise, when $\kappa \gg 1$.
\begin{figure}
\centering
\includegraphics[width=1.0\linewidth]{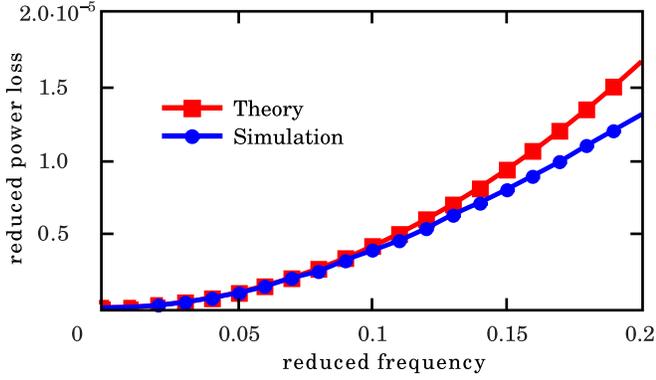}
\caption{(Color online) The comparison of the analytical and numerical results while the circularly-polarized field is applied. The value of the relationship between the magnetic and thermal energies is chosen as $\kappa = 5$, the value of the field amplitude is chosen as $h_{m} = 0.01$. The theory gives larger values, and the difference increases with frequency}
\label{fig:onepart_res_cp_theor}
\end{figure}
\begin{figure}
\centering
\includegraphics[width=1.0\linewidth]{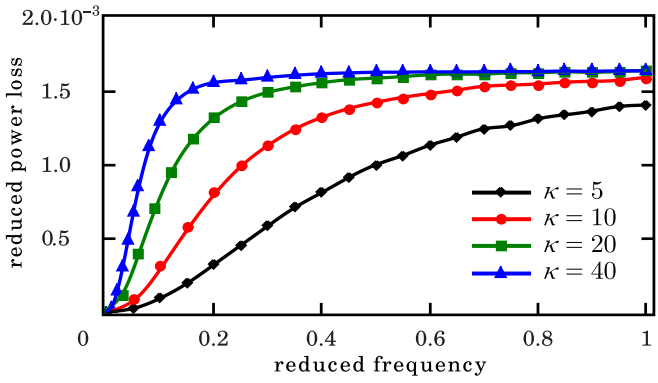}
\caption{(Color online) The frequency dependencies of the power loss for different values of the relationship between the magnetic and thermal energies while the circularly-polarized field is applied. The value of the field amplitude is chosen as $h_{m} = 0.05$. The behaviour is similar to the case of the linearly-polarized field action (see Fig.~\ref{fig:onepart_res_lp_num}), but the values are approximately two times larger}
\label{fig:onepart_res_cp_num}
\end{figure}
To verify our analytical findings and see the domain of applicability of our results, the set of simulation runs has been performed on the basis of the numerical integration of Eq.~(\ref{eq:Langevin_angular_eff_red}). As we can see from  Fig.~\ref{fig:onepart_res_cp_theor}, the analytical and numerical results correlate well for practically interesting values of the noise intensities and external field parameters. Thus, for $\kappa = 5$ and $h_{m} = 0.01$, Eq.~(\ref{eq:Q_stoch_cp}) gives feasible results for the field frequencies up to $\widetilde{\Omega} \sim 0.1$. The study of the power loss in the whole acceptable range of parameters can be performed only numerically. As expected, when the ratio of the magnetic and thermal energies $\kappa$ grows, the power loss tends to the noiseless limit values. As seen from Fig.~\ref{fig:onepart_res_cp_theor}, for small frequencies the difference between the power loss values grows nonlinearly on $\kappa$. However, this difference decreases on $\widetilde{\Omega}$ and is small enough for $\widetilde{\Omega} \sim 1$. At high frequencies $\widetilde{Q}$ tends to a constant, which is proportional to $h_{m}$ in a wide range of $\kappa$. Finally, comparing Fig.~\ref{fig:onepart_res_cp_num} and Fig.~\ref{fig:onepart_res_lp_num} one can conclude that the dependencies obtained for the circularly- and linearly-polarized fields are qualitatively similar. In accordance with the noiseless results Eq.~(\ref{eq:Q_small_osc}), the high-frequency limit value for the circularly-polarized field is approximately two times larger than for the linearly-polarized one. Here, the two-fold difference is actual for the entire frequency range that is a consequence of the thermal bath presence and is not typical for the noiseless case. The integrated results for different $\kappa$ and $h_{m}$ were illustrated in Fig.~\ref{fig:onepart_res_cp_num_3d}, where all mentioned trends and features are shown in one set. As seen from Fig.~\ref{fig:onepart_res_cp_num_3d}a, for small frequencies ($\widetilde{\Omega} = 0.05$ in the figure) the character of the dependencies is in a good agreement with Eq.~(\ref{eq:Q_stoch_cp}), where the power loss is proportional to $\kappa^2$ and $h_{m}^2$. In contrast, as follows from Fig.~\ref{fig:onepart_res_cp_num_3d}b, for large frequencies ($\widetilde{\Omega} = 1$ in the figure) the dependence on $\kappa$ becomes weak enough, while the nonlinear increase on $h_{m}$ remains.
\begin{figure}
\centering
\includegraphics[width=1.0\linewidth]{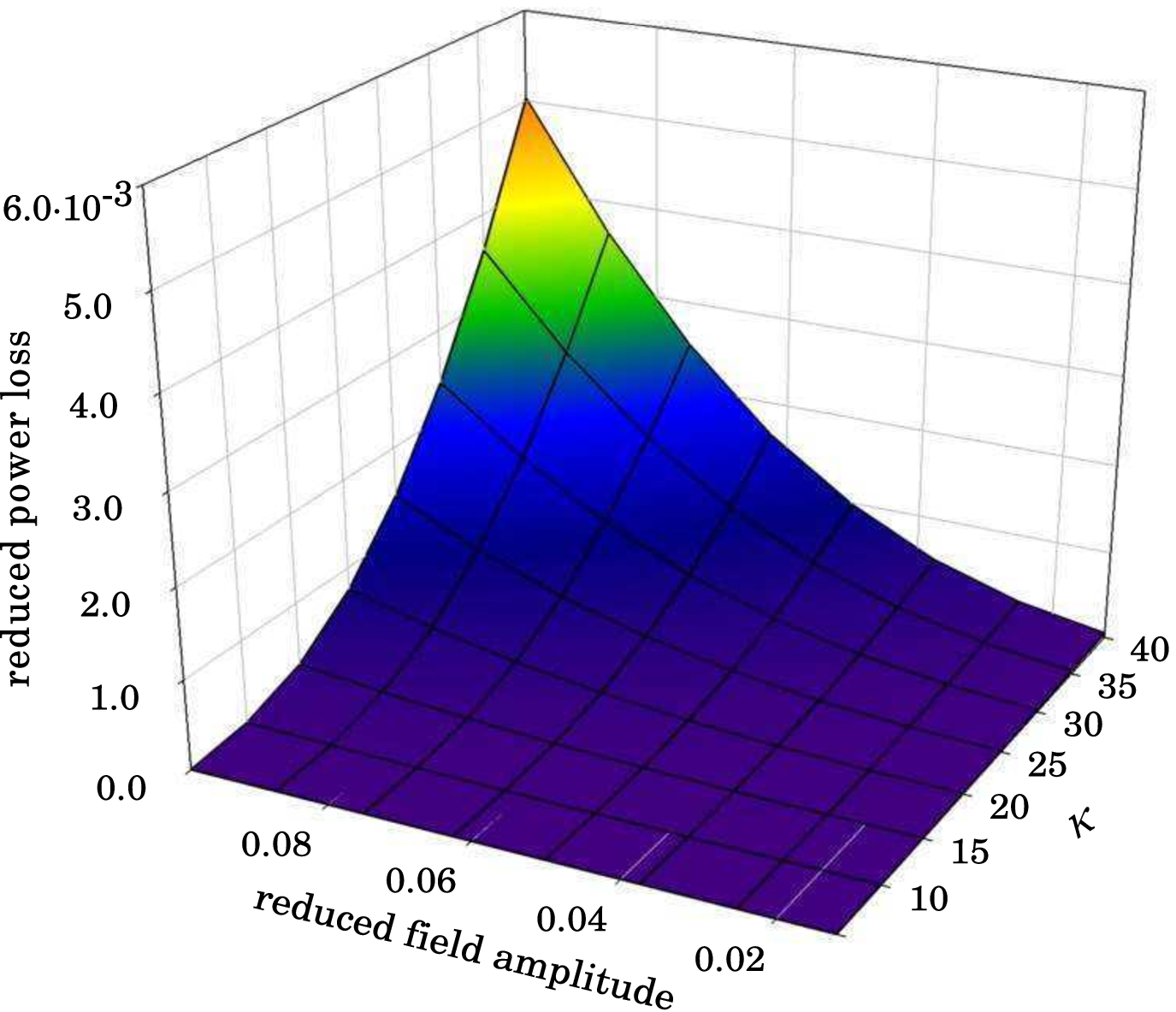}
\includegraphics[width=1.0\linewidth]{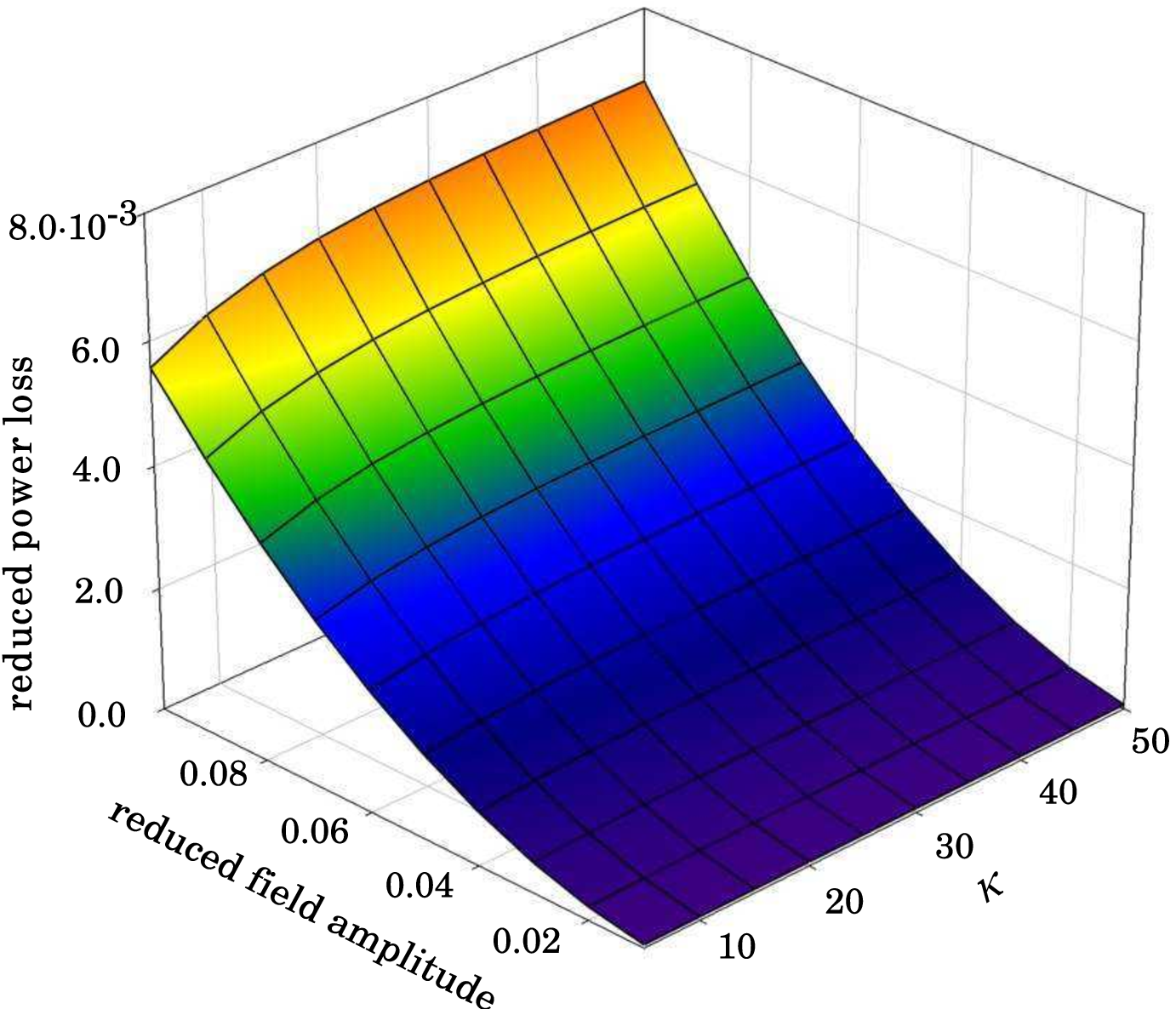}
\caption{(Color online) The dependencies of the power loss on the field frequency and the relationship between the magnetic and thermal energies while the circularly-polarized field is applied. Plot a) demonstrates a strong nonlinear decrease in the power loss on both parameters, when the frequency is low (the value of the field frequency is chosen as $\widetilde{\Omega} = 0.05$). Plot b) demonstrates a weak dependence of the power loss on the relationship between the magnetic and thermal energies, when the frequency is high (the value of the field frequency is chosen as $\widetilde{\Omega} = 1$)}
\label{fig:onepart_res_cp_num_3d}
\end{figure}

As a preliminary summary, we want to underline the following. Thermal fluctuations lead to decay of the power loss, but the character of the dependencies $\widetilde{Q}(\widetilde{\Omega})$ remains similar to the noiseless limit. The difference caused by thermal noise decays on the field frequency. The results presented for the circularly-polarized field correlate qualitatively with the results obtained in \cite{Raikher2011, PhysRevE.83.021401}, while for the linearly-polarized field we, in fact, briefly repeated the results presented in \cite{0953-8984-15-23-313}. But we conducted our investigation for all these cases in a uniform and rigorous manner, which is based on the Fokker-Planck equation Eq.~(\ref{eq:FP}) and power loss determination Eq.~(\ref{eq:def_Q}) for analytical estimations. In turn, the explicit form of the Langevin equation Eq.~(\ref{eq:Langevin_angular_eff_red}) is used for the numerical simulation, which gives us a deep understanding of the role of thermal fluctuations in absorption of an alternating field by rigid dipoles.

\subsection{Influence of the inter-particle interaction}

The inter-particle interaction can essentially impact the response to an external field, and accounting of this interaction is important in a view of real ferrofluid applications. Even for relatively small volume fractions (for example, 1 $\%$), due to the repulsion caused by the surfactant covering of each particle and the long-range dipole interaction between particles, the behaviour of each particle will be different from the single-particle approximation outlined above. The dipole interaction intends to join particles into clusters. Such structures are extremely undesirable for hyperthermia by reason of further metabolism and excretion. To prevent this clustering, the particles are coated with a surfactant providing repulsion. Competition between the above mentioned interactions can modify the specific power loss of each particle in a wide range, that is in focus of this section.

The inter-particle interaction, by and large, increases the magnetic energy, and there are two consequences of this fact. Firstly, the regular component of motion becomes strong due to the interaction, and the stochastic component, in contrast, is suppressed. At a glance, such suppression can result in the power loss increase. Secondly, the interaction fixes the particle magnetic moments, and the response to an external field becomes poor. This trend has led to the power loss decrease that was observed in recent experiments \cite{0022-3727-49-29-295001, doi:10.1109/TMAG.2016.2516645, doi:10.1063/1.4974803, Arteagacardona2016636, doi:10.1021/acs.molpharmaceut.5b00866, doi:10.1021/acsnano.7b01762, doi:10.2147/IJN.S141072, doi:10.1166/sam.2017.2948, doi:10.1021/acsnano.7b01762}. Nevertheless, the mentioned experiments do not describe all possible situations and inspire future experimental and numerical investigations.

As stated above, the interaction leads to the cluster formation, when strong dipole fields hold each particle that obstructs further translational and rotational motions. At the same time, the clusters are an origin of a few phenomena, which modify essentially the power loss value. Firstly, each particle tries to reduce its energy and get the equilibrium or quasi-equilibrium state caused by the interaction. In particular, such states are generated by the dipole field, which tries to align the particle magnetic moments along the defined directions. Due to thermal fluctuations, the magnetic moment together with the particle can perform the transition or switching between these states. Such switchings change the magnetic alignment in the clusters and lead to frustrations. Under some circumstances, this can result in the power loss increase. Secondly, large enough fluctuations can destroy the clusters completely and the particle response to an external field becomes better. Here, the power loss also increases, and we interpret this as the constructive role of thermal noise. We have studied in-depth all the above phenomena including the influence of the system parameters on their conditions of occurrence. In this regard, the volume fraction, noise intensity, and surfactant characterizations are the most interesting.

To explain the mentioned phenomena, let us consider the mechanisms of cluster formation in detail. From a position of the minimal magnetostatic energy, the particles should be closer to each other. Also, the magnetic moment of each particle should be directed along the resulting dipole field, which is generated by other particles. Since the magnetic lines of force are closed curves, two trends take place. Firstly, the particle magnetic moments try to be aligned along one direction, and this leads to the chain-like cluster formation. Secondly, the chain fragments tend to be arranged in the antiparallel way and attract each other forming the antiferromagnetic structure. To prevent such agglomeration, each particle is covered with a special surfactant, which provides steric repulsion. The competition between the dipole attraction and steric repulsion can lead to quite different results. We need to underline that since the magnitude of magnetization is important for the performance of the hyperthermia method, it is reasonable to synthesize particles with magnetization as large as possible. As a consequence, the intensity of the dipole interaction should increase, and the clusters will become denser. Therefore, the actuality of interaction accounting will become larger.
\begin{figure}
\centering
\includegraphics[width=0.9\linewidth]{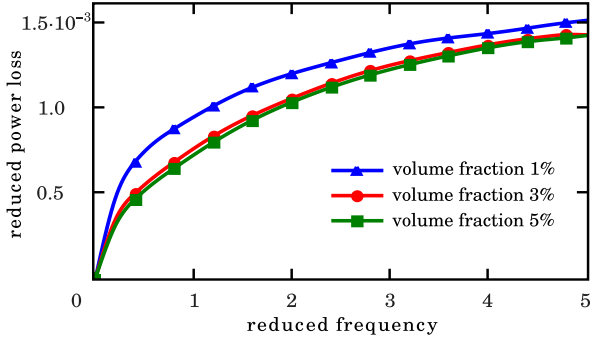}
\caption{(Color online) Ensemble simulation results: the frequency dependencies of the power loss for different values of the particle volume fractions while the circularly-polarized field is applied. The value of the field amplitude is chosen as $h_{m} = 0.05$, the value of the relationship between the magnetic and thermal energies is chosen as $\kappa = 10$, the value of the potential barrier depth is chosen as $\varepsilon = 0.04$, the dimensionless equilibrium distance between two particles is chosen as $\sigma = 2.25 $. The particles are aggregated into chine-like structures}
\label{fig:interact_vol_frac_nonclust}
\end{figure}

\begin{figure}
\centering
\includegraphics[width=0.9\linewidth]{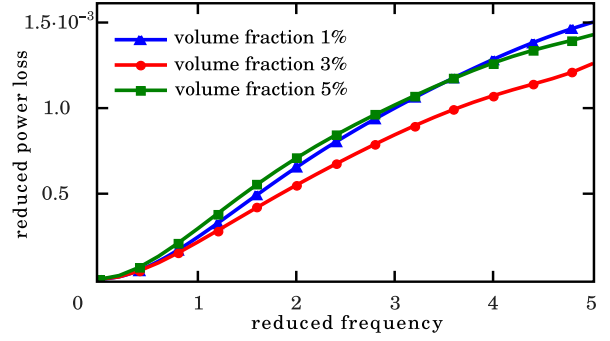}
\caption{(Color online) Ensemble simulation results: the frequency dependencies of the power loss for different values of the particle volume fractions while the circularly-polarized field is applied. The value of the field amplitude is chosen as $h_{m} = 0.05$, the value of the relationship between the magnetic and thermal energies is chosen as $\kappa = 25$, the value of the potential barrier depth is chosen as $\varepsilon = 0.04$, the dimensionless equilibrium distance between two particles is chosen as $\sigma = 2.25 $. The particles are aggregated into dense clusters, consists of a chain fragments}
\label{fig:interact_vol_frac}
\end{figure}

Various system parameters influence the cluster formation process differently, and this impact can be ambiguous. As a rule, increase in the volume fraction leads to the particles agglomeration that, in turn, results in the power loss decrease (see Fig.~\ref{fig:interact_vol_frac_nonclust}). We tend to explain this by different cluster types. When the volume fraction is small, the chain-like structures are formed. The particles in such structures interact weakly, therefore, they are more sensitive to an external field. For a larger volume fraction, the short chain fragments join each other forming denser structures with stronger interaction. And these aggregated structures have a weak response to an external field. But there are some exceptions from this trend. Firstly, when the noise intensity is small enough, the role of the particle concentration can be negligible. This happens because the formed clusters for different volume fractions have the similar structure and remain stable. Secondly, the power loss can increase with the volume fraction, as seen from Fig.~\ref{fig:interact_vol_frac} by the following reasons. When the inter-particle interaction in the clusters is strong, remagnetization of the whole cluster requires a stronger field, and, as a consequence, the hysteresis loop widens. It is confirmed by the fact that the curve for 5$\%$ in Fig.~\ref{fig:interact_vol_frac} exceeds the curve for 3$\%$. This effect disappears for larger frequencies, when the clusters do not reverse completely during the field period. Here, strong interaction suppresses the response of each particle, and, in contrary, the power loss becomes smaller in comparison with the case of lower volume fractions. The ambiguous character of the power loss dependence on the particles concentration has an experimental confirmation n the measurement of specific absorbtion rate of the iron oxide nanoparticles, dispersed in agar \cite{doi:10.1007/s11051-015-2921-9} (See Fig.~4, page 3).

\begin{figure}
\centering
\includegraphics[width=0.9\linewidth]{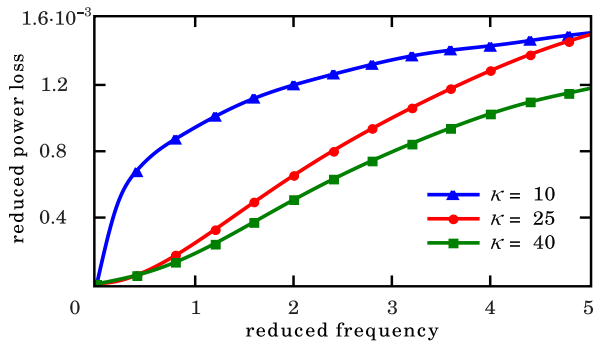}
\caption{(Color online) Ensemble simulation results: the frequency dependencies of the power loss for different values of the relationship between the magnetic and thermal energies while the circularly-polarized field is applied. The value of the field amplitude is chosen as $h_{m} = 0.05$, the value of the volume fraction is chosen as 1$\%$, the value of the potential barrier depth is chosen as $\varepsilon = 0.04$, the dimensionless equilibrium distance between two particles is chosen as $\sigma = 2.25 $. The origin of the unusual curve behaviour for $\kappa = 25$ consists in the particles switchings between the quasi-equilibrium states, which are constituted by the local dipole fields}
\label{fig:interact_switch_clust}
\end{figure}

Thermal fluctuations break the order with increasing temperature, and this can affect on the ferrofluid response differently. A very interesting effect occurs under some circumstances, when the magnetic energy of a particle is larger than the thermal one, but not so large to exclude the essential fluctuations during the field period. In this case, each particle in a cluster is in the quasi-equilibrium state provided by the resulting dipole field. The particle magnetic moment fluctuates most of the time in the vicinity of one of these states. Due to rare, but strong fluctuations, the particle can transit from one state to another. Such phenomenon is similar to the relaxation of the magnetic moment in the fixed uniaxial particle described in \cite{PhysRev.130.1677} or to the field-induced switching in the same particle considered in \cite{PhysRevLett.97.227202}. The transition process proceeds fast enough, but during it each particle is frustrated and characterized by a high energy in an external field. Generally, this leads to the power loss increase, especially for high frequencies, when the time of one transition becomes comparable with the field period. Such increment of the energy dissipation requires a number of conditions, since a lot of factors impact the ratio between the thermal and deterministic energies, namely, noise intensity, surfactant parameters, and cluster types. In Fig.~\ref{fig:interact_switch_clust}, the power loss increase by thermal fluctuations is reflected in the behaviour of the curve for $\kappa = 25$. As seen, the curves for $\kappa = 25$ and $\kappa = 40$ coincide for small frequencies. But while the frequency increases, the difference between the curves for $\kappa = 25$ and $\kappa = 40$ becomes larger. As expected, for frequencies more than $\widetilde{\Omega} = 5$ the power loss values for $\kappa = 25$ will be larger even than for $\kappa = 10$. In this sense, this effect can be more productive than the cluster decomposition described below.

\begin{figure}
\centering
\includegraphics[width=0.9\linewidth]{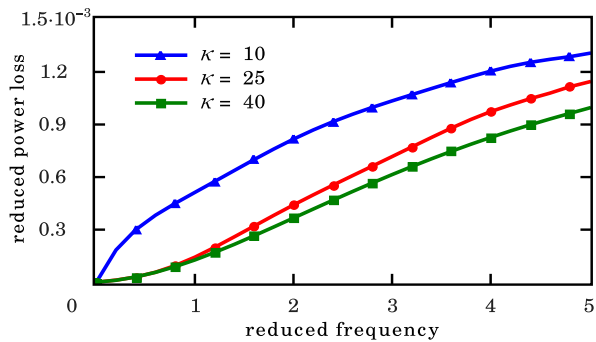}
\caption{(Color online) Ensemble simulation results: the frequency dependencies of the power loss for different values of the relationship between the magnetic and thermal energies while the circularly-polarized field is applied. The value of the field amplitude is chosen as $h_{m} = 0.05$, the value of the volume fraction is chosen as 3$\%$, the value of the potential barrier depth is chosen as $\varepsilon = 0.02$, the dimensionless equilibrium distance between two particles is chosen as $\sigma = 2.25 $. The curves positions suggest that the clusters are broken completely, when the noise intensity increases}
\label{fig:interact_construct_noise_role}
\end{figure}

Despite the noise suppresses the response of each particle to an external field, for the interacting ensemble it can lead to quite different results. The particles in dense clusters are strongly bonded and weakly exposed to an external field. But if the thermal energy is comparable with the magnetic one, thermal fluctuations make the particles free that completely prevents the cluster formation. In this way, thermal fluctuations increase the particle response to an external periodic field and, correspondingly, the energy absorbed from this field. We tend to interpret this as the constructive role of noise. The results of the set of simulations confirming this phenomenon are depicted in Fig.~\ref{fig:interact_construct_noise_role}. Firstly, for large noise intensities ($\kappa = 10$) the particle aggregation does not occur, and the phenomenon is extremely pronounced. Secondly, for smaller noise intensities the clusters, nevertheless, are partially formed, but they can be destructed depending on other parameters. As seen, at low frequencies the curve for $\kappa = 25$ almost coincides with the curve for $\kappa = 40$, and both of them are located far above the curve for $\kappa = 10$. In both these cases, the similar clusters are formed, and they are not completely broken by thermal noise. But for high frequencies, when the field-induced oscillations of the particles promote the destruction of clusters, the differences between the mentioned curves increases.
In summary, we point out that the inter-particle interaction as well as thermal noise suppresses the particle response to an external field. At the same time, both these factors can compete with each other, and the resulting power loss in a periodic external field can vary widely. The cluster formation due to the dipole interaction with further cluster transformations and destruction due to thermal bath and external periodic field actions plays the crucial role here. The main practical result is that the temperature increase promotes the energy absorbtion. There are two reasons for this: the switchings of the particles in clusters due to fluctuations that temporary frustrates the magnetic order, and the complete decomposition of the clusters by noise that increases the particle response to an external field.

\section{Conclusions}

We present the comprehensive study of the interaction of identical spherical uniform particles of magnetization $M$, which are placed in a fluid of viscosity $\eta$, with a time-periodic external magnetic field of amplitude $H_m$ and frequency $\Omega$. We imply that the magnetization is locked in the crystal lattice by high magnetic anisotropy. This approach is also the so-called rigid dipole model. The main efforts are devoted to the environmental heating that is a result of the interaction of the external field with the particles. We characterize this heating by the dimensionless power loss $\widetilde{Q}$. The investigation is conducted in three stages. Firstly, we have performed the consideration of the single-particle noise-free or dynamical approximation. Secondly, we have accounted both the external field and thermal bath using the Langevin and Fokker-Planck equations. Finally, the inter-particle interaction effects have been investigated numerically using the Barnes-Hut algorithm \cite{Barnes-Hut-Nature1986} and CUDA technology \cite{Sanders2011CUDA}. We summarize our findings as follows.

In the dynamical approximation, the dependencies $\widetilde{Q}(\widetilde{\Omega})$ at $\widetilde{\Omega} \ll 1$ are quite different depending on the external field polarization type. This occurs because the rotational trajectories of the particle and its response to the external fields with various polarizations are different enough for low frequencies. Thus, $\widetilde{Q} = \widetilde{\Omega}^2$ (the dependence on the field amplitude is absent) for the circularly-polarized field and $\widetilde{Q} \sim \widetilde{\Omega} h_{m}$ for the linearly-polarized one. This is a purely dynamical effect, and since it is not observed in the presence of thermal bath, there is no any reason to speak about its practical meaning.

The dependencies $\widetilde{Q}(\widetilde{\Omega})$ have a saturated character and at $\widetilde{\Omega} \ll 1$ tend to the constant, which is proportional to $h_{m}^{2}$. The values of $\widetilde{Q}$ for the circularly-polarized field are, at least, two times bigger than for the linearly-polarized one. In contrast to the low-frequency case, the main trends here remain, when thermal fluctuations are present. Therefore, the results of the dynamical approximation for high frequencies of the external field can be applicable to a real particle in a fluid.

Thermal fluctuations reduce the power loss in a nonlinear way at small frequencies. Thus, for the circularly-polarized external field this reduction is realized as $\kappa^2$ ($\kappa$ is the relationship between the magnetic and thermal energies), when the conditions $\kappa h_{m} \ll 1$, $\kappa \widetilde{\Omega} \ll 1$ hold. At the same time, for large enough frequencies ($\widetilde{\Omega} \sim 1$), the difference between the power loss values at various noise intensities, which is defined by $\kappa$, becomes less pronounced. At last, when $\widetilde{\Omega} \ll 1$, the results are almost indistinguishable from the dynamical approximation in a wide range of $\kappa$. This behaviour needs to be accounted to improve the performance of the magnetic fluid hyperthermia method.

The power loss values for the circularly-polarized field approximately two times exceed the values for the linearly-polarized one in the whole frequency range. This is obvious for the quasi-equilibrium approximation described in \cite{Rosensweig2002370} and follows from our analysis of the noise-free limit for high frequencies. But it is rather unexpected in a light of low-frequency behaviour in the dynamical approximation described above. This fact suggests that thermal fluctuations change qualitatively the particle behaviour.

The interaction between the particles has a strong impact on the ensemble susceptibility to the external periodic fields. Because of the cluster formation, each particle is under the action of the strong enough resulting dipole field that complicates its response. Evidently, this results in a considerable decay of the power loss that is in agreement with the experiments. At the same time, the different cluster types are possible, and even small changes of the field, particle or fluid parameters can cause the quite different structure of the ensemble. These effects are especially actual for low frequencies, when the clusters inverse their magnetization completely during the field period. This leads to the strong difference between the $\widetilde{Q}$ values in comparison with the results of the single-particle case. This difference reduces with the field frequency, since for high frequencies each particle performs oscillations around its initial position without full inversion of the magnetization. Therefore, the interaction impact is not critical, when $\widetilde{\Omega} \gg 1$, excluding the specific situation described below.

However, the inter-particle interaction and thermal noise are the competing factors. We have revealed two situations, when thermal noise leads to the power loss increase in the presence of inter-particle interaction. We associate these phenomena with the constructive role of noise. In both mentioned cases, the resulting values of $\widetilde{Q}$ can be almost equal to the value for the single-particle approximation. The first of them occurs, when the noise intensity is large. The clusters are destroyed here that results in the $\widetilde{Q}$ increase as a consequence of the better response of the particles to the external field. And under certain other conditions, larger noise intensity corresponds to larger power loss values. The second of them occurs, when the noise intensity is not so large, and the magnetic energy is comparable with the thermal one. Here, the fluctuations partially blur the particles order in the clusters that also leads to the significant increase in $\widetilde{Q}$ for high field frequencies. We have explained this by rare switchings of the particles in the clusters between the quasi-equilibrium states formed by the resulting dipole fields. The switching process is performed through the frustrated states characterised by high energies. Exactly the high energy consumption in these mentioned states causes the observed increase in the power loss.

\section*{ACKNOWLEDGMENTS}

The authors express appreciation to S.~I.~Denisov and Yu.~S.~Bystrik for the valuable comments and discussion. The authors are also grateful to the Ministry of Education and Science of Ukraine for partial financial support under Grant No.~0116U002622.

\section*{References}

\bibliography{Lyutyy_Ukraine_PowerLoss_FNP_Ensemble_Revised}

\end{document}